\documentclass[a4paper,11pt]{article}
\usepackage{jcappub} 
\usepackage{lineno}
\usepackage{natbib}
\usepackage[T1]{fontenc}
\usepackage{cleveref}
\usepackage[normalem]{ulem}

\newcommand\aj{\ref@jnl{AJ}}
\newcommand\psj{\ref@jnl{PSJ}}
\newcommand\araa{\ref@jnl{ARA\&A}}
\newcommand\apj{\ref@jnl{ApJ}}
\newcommand\apjl{\ref@jnl{ApJL}}     
\newcommand\apjs{\ref@jnl{ApJS}}
\newcommand\ao{\ref@jnl{ApOpt}}
\newcommand\apss{\ref@jnl{Ap\&SS}}
\newcommand\aap{\ref@jnl{A\&A}}
\newcommand\aapr{\ref@jnl{A\&A~Rv}}
\newcommand\aaps{\ref@jnl{A\&AS}}
\newcommand\azh{\ref@jnl{AZh}}
\newcommand\baas{\ref@jnl{BAAS}}
\newcommand\icarus{\ref@jnl{Icarus}}
\newcommand\jaavso{\ref@jnl{JAAVSO}}  
\newcommand\jrasc{\ref@jnl{JRASC}}
\newcommand\memras{\ref@jnl{MmRAS}}
\newcommand\mnras{\ref@jnl{MNRAS}}
\newcommand\pra{\ref@jnl{PhRvA}}
\newcommand\prb{\ref@jnl{PhRvB}}
\newcommand\prc{\ref@jnl{PhRvC}}
\newcommand\prd{\ref@jnl{PhRvD}}
\newcommand\pre{\ref@jnl{PhRvE}}
\newcommand\prl{\ref@jnl{PhRvL}}
\newcommand\pasp{\ref@jnl{PASP}}
\newcommand\pasj{\ref@jnl{PASJ}}
\newcommand\qjras{\ref@jnl{QJRAS}}
\newcommand\skytel{\ref@jnl{S\&T}}
\newcommand\solphys{\ref@jnl{SoPh}}
\newcommand\sovast{\ref@jnl{Soviet~Ast.}}
\newcommand\ssr{\ref@jnl{SSRv}}
\newcommand\zap{\ref@jnl{ZA}}
\newcommand\nat{\ref@jnl{Nature}}
\newcommand\iaucirc{\ref@jnl{IAUC}}
\newcommand\aplett{\ref@jnl{Astrophys.~Lett.}}
\newcommand\apspr{\ref@jnl{Astrophys.~Space~Phys.~Res.}}
\newcommand\bain{\ref@jnl{BAN}}
\newcommand\fcp{\ref@jnl{FCPh}}
\newcommand\gca{\ref@jnl{GeoCoA}}
\newcommand\grl{\ref@jnl{Geophys.~Res.~Lett.}}
\newcommand\jcp{\ref@jnl{JChPh}}
\newcommand\jgr{\ref@jnl{J.~Geophys.~Res.}}
\newcommand\jqsrt{\ref@jnl{JQSRT}}
\newcommand\memsai{\ref@jnl{MmSAI}}
\newcommand\nphysa{\ref@jnl{NuPhA}}
\newcommand\physrep{\ref@jnl{PhR}}
\newcommand\physscr{\ref@jnl{PhyS}}
\newcommand\planss{\ref@jnl{Planet.~Space~Sci.}}
\newcommand\procspie{\ref@jnl{Proc.~SPIE}}

\newcommand\actaa{\ref@jnl{AcA}}
\newcommand\caa{\ref@jnl{ChA\&A}}
\newcommand\cjaa{\ref@jnl{ChJA\&A}}
\newcommand\jcap{\ref@jnl{JCAP}}
\newcommand\na{\ref@jnl{NewA}}
\newcommand\nar{\ref@jnl{NewAR}}
\newcommand\pasa{\ref@jnl{PASA}}
\newcommand\rmxaa{\ref@jnl{RMxAA}}
\usepackage{xcolor}
\definecolor{seagreen}{rgb}{0.05, 0.65, 0.20}

\title{Quantum Tunneling of Primordial Black Holes to White Holes:\\ Rates, Constraints, and Implications for Fast Radio Bursts}
\author{Christopher Ewasiuk}
\emailAdd{cewasiuk@ucsc.edu}
\author{and Stefano Profumo}
\affiliation{Santa Cruz Institute for Particle Physics and}
\affiliation{Department of Physics, University of California, Santa Cruz \\Santa Cruz, CA, 95064, USA}
\emailAdd{profumo@ucsc.edu}

\abstract{We calculate the present-day and cosmological volumetric rate of primordial black hole (PBH) quantum tunneling events to white holes, incorporating the competition between Hawking evaporation and tunneling, cosmological depletion, realistic mass-dependent abundance constraints, extended mass functions, and the alternative memory-burden scenario.  The burst rate is maximized along a narrow ridge in the mass--tunneling-parameter plane where the effective PBH lifetime is comparable to the age of the Universe.  Within the canonical Planck-star range of tunneling timescales, FRB-level rates arise only in two highly restricted regions: a low-mass window near the evaporation boundary, and a narrow sequential window where evaporation precedes tunneling; broadening the PBH mass function does not qualitatively alter this conclusion.  We further assess observational constraints from FRB repetition statistics, radio spectral properties, prompt and diffuse gamma-ray limits, host-galaxy demographics, and gravitational-wave signatures.  All current observations are consistent with a subdominant white-hole contribution to the FRB population but strongly disfavor a dominant origin.  The rate calculation does not generically support FRB-level event densities, and any viable FRB interpretation requires narrow, fine-tuned, and strongly assumption-dependent corners of parameter space.
}

\begin{document}

\maketitle
\flushbottom

\section{Introduction}
\label{sec:intro}

Black holes are traditionally considered long-lived classical objects, whose only quantum effect is slow evaporation via Hawking radiation on timescales vastly exceeding the age of the Universe for astrophysical masses \cite{Hawking:1974rv}. However, several approaches to quantum gravity suggest that black holes may undergo a quantum transition to white hole (WH) states, a process sometimes referred to as a ``Planck-star bounce'' \cite{Haggard:2014rza,Rovelli:2014cta,Christodoulou:2016vny}. In this picture, the classical singularity is replaced by a region of finite density that eventually tunnels to a white hole geometry, releasing the remaining mass-energy in a burst of high-energy radiation.

If primordial black holes (PBHs) are formed in the early Universe \cite{Carr:1974nx, Carr:2020xqk} and if they make up a fraction $f_{\rm PBH}$ of the dark matter density, their quantum tunneling decays could be an observable probe of quantum gravity. The volumetric burst rate $R_{\rm PBH}$ depends on both their abundance and the quantum tunneling lifetime \cite{Haggard:2014rza,Rovelli:2014cta,Christodoulou:2016vny}:
\begin{equation}
    \tau(M) = \alpha\bigg(\frac{M}{M_{\rm Pl}}\bigg)^2 t_{\rm Pl},
    \label{eq:intro_tau}
\end{equation}
where $\alpha$ is a dimensionless parameter encoding non-perturbative gravitational dynamics. In current heuristic models, $\alpha$ is assumed to be ${\cal O}(0.01$–$1$), yielding relatively prompt bounces for light PBHs.

Fast radio bursts (FRBs) \cite{CHIMEFRB:2021srp}, bright millisecond-duration radio transients of unknown origin, have motivated interest in PBH quantum bursts as possible progenitors \cite{Barrau:2014yka,Barrau:2015uca}. If even a small fraction of PBHs exploded today, their integrated rate could approach the observed FRB volumetric density of $10^{4-5}$~Gpc$^{-3}$~yr$^{-1}$. However, existing estimates of $R_{\rm PBH}$ are highly uncertain, with only order-of-magnitude estimates available and little exploration of the full $(M,\alpha)$ parameter space, nor of the depletion of PBHs that would have tunneled earlier in cosmic history.

In this work, we revisit the calculation of $R_{\rm PBH}$ in detail. We:
\begin{enumerate}
    \item Derive the expected volumetric rate as a function of $M$, $\alpha$, and $f_{\rm PBH}$, accounting for cosmological depletion of early tunneling events;
    \item Perform a comprehensive parameter scan over $\alpha\in[10^{-4},10^{25}]$ and $M\in[10^{8},10^{20}]$~kg, producing updated constraints and rate predictions;
    \item Compare the results to the observed FRB rate and assess detectability with current instruments; and 
    \item Discuss the theoretical implications of requiring $\alpha \gg 1$ for FRB-level rates and the prospects for probing quantum gravity with PBH tunneling.
\end{enumerate}

\noindent Relative to earlier order-of-magnitude estimates~\cite{Barrau:2014yka,Barrau:2015uca}, the principal novelties of our treatment are: (a)~explicit inclusion of the competition between Hawking evaporation and quantum tunneling through a combined effective lifetime; (b)~cosmological depletion of PBHs that tunneled at earlier epochs; (c)~incorporation of realistic, mass-dependent upper bounds $f_{\rm PBH,max}(M)$ from microlensing, CMB, and evaporation constraints; (d)~a systematic scan of the full $(M,\alpha)$ parameter space; (e)~extension to lognormal mass functions; and (f)~a parallel analysis under the memory-burden hypothesis, which enforces sequential rather than competitive evolution.

Our findings show that the present-day PBH tunneling burst rate is sharply peaked
along the locus where the effective lifetime satisfies $\tau_{\rm eff}(M) \simeq t_0$, with $t_0$ the current age of the universe, 
producing a narrow ridge on the $(M,\alpha)$ plane.
For the canonical Planck-star range $10^{-2} \lesssim \alpha \lesssim 1$,
FRB-level volumetric rates,
$R \sim 10^{4-5}\,\mathrm{Gpc}^{-3}\,\mathrm{yr}^{-1}$,
are not generically obtained but may arise  in two highly-restricted regions:

(i) a low-mass window near the evaporation boundary
$M \sim 10^{11-12}\,\mathrm{kg}$,
where tunneling becomes competitive with Hawking depletion,
and 

(ii) a narrow sequential window in the memory-burden scenario,
where evaporation first reduces the mass before tunneling activates.

Outside these finely tuned regions,
the rate is exponentially suppressed or too small to be observationally relevant.
Including realistic PBH abundance limits and cosmological evolution
further compresses the viable parameter space,
so that PBH tunneling can at most account for a subpopulation of FRBs
under highly specific mass and abundance assumptions.

The remainder of our study is organized as follows. In Sec.~\ref{sec:theory} we review the
black-to-white hole tunneling framework, including its quantum-gravity
motivation, lifetime scaling, and emission signatures. Sec.~\ref{sec:detectability}
discusses the multi-channel detectability of Planck-star bursts (PSBs) across radio,
gamma-ray, and cosmic-ray observations. Sec.~\ref{sec:FRB} reviews observational status and constraints on FRB rates. In Sec.~\ref{sec:rate} we derive the
expected volumetric burst rate from black-to-white hole transition, present results for the
monochromatic mass function and present-day rate, discuss  redshift evolution, extend the analysis
to non-monochromatic mass functions, and examine the
theoretical status of the tunneling parameter $\alpha$. Sec.~\ref{sec:obs_gamma} studies gamma-ray and multiwavelength signatures and constraints. Finally, we synthesize
the implications for FRB phenomenology and quantum gravity in
Sec.~\ref{sec:conclusions}.

\section{Black-to-White Hole Tunneling Model}
\label{sec:theory}


In general relativity, the gravitational collapse of matter beyond its Schwarzschild radius leads to the formation of an event horizon enclosing a spacetime singularity. This description is widely believed to be incomplete, since spacetime singularities signal the breakdown of the classical gravitational theory. Loop Quantum Gravity (LQG) provides a background-independent, non-perturbative quantization of spacetime geometry that naturally resolves singularities through quantum gravitational repulsion at Planckian densities \cite{Ashtekar:2006rx,Rovelli:2004tv}.
In this context, black holes are conjectured to be metastable states that can undergo a quantum tunneling process into white holes \cite{Haggard:2014rza,Christodoulou:2016vny}. The transition connects the trapped region inside the horizon with an expanding geometry, enabling information and matter initially captured by the black hole to be released after a finite delay. This is the basis for the so-called ``Planck star'' scenario \cite{Rovelli:2014cta}, in which the black hole core never reaches a singularity but bounces due to quantum pressure, subsequently emerging as a white hole.

\subsection{Lifetime of the Metastable Black Hole}

The quantum tunneling process is not instantaneous. Calculations in simplified models and spinfoam quantizations of LQG indicate that the expected lifetime of a black hole before transitioning to a white hole scales quadratically with its mass; in Planck units $(G = c = \hbar = 1)$, Eq.~\eqref{eq:intro_tau} reads:
\begin{equation}
    \tau(M) = \alpha\,M^2\,,
    \label{eq:tau}
\end{equation}
where $\alpha$ is a dimensionless constant that encodes details of the transition amplitude \cite{Christodoulou:2016vny}. This scaling differs from the classical Hawking evaporation timescale $\tau_{\rm Hawking}\propto M^3$, potentially allowing much heavier black holes than the Hawking limit ($M\sim10^{12}$~kg) to explode in the present epoch. 
Phenomenological estimates place $\alpha$ in the range $10^{-2}\lesssim\alpha\lesssim10^{-1}$ \cite{Christodoulou:2016vny,Barrau:2014fcg}, although the value of $\alpha$ is highly uncertain. With these values, primordial black holes (PBHs) of mass $M\sim10^{22}$--$10^{24}$~kg, formed in the early Universe, would have tunneling times comparable to the age of the Universe $t_0\approx 4.3\times10^{17}$~s and could transition to white holes today.

\subsection{Emission Signatures}

The black-to-white hole tunneling event is expected to release a significant fraction of the trapped energy of the PBH via two main channels:
\begin{itemize}
    \item A \textbf{low-frequency coherent burst}, with a characteristic frequency set by the Schwar\-zschild radius of the exploding PBH:
    \begin{equation}
        \nu_{\rm radio} \sim \frac{c}{2\pi R_S} = \frac{c^3}{4\pi G M} \approx 300~\text{GHz}\left(\frac{10^{23}~\text{kg}}{M}\right).
        \label{eq:nu_radio}
    \end{equation}
    Thus, a signal at GHz frequencies or below requires $M\lesssim10^{25}$~kg; however, secondary emission mechanisms---such as coherent plasma processes in the white hole outflow---would shift the characteristic frequency downward. As such, lighter black holes are also in principle possible counterparts to FRBs. We discuss this in detail in Sec.~\ref{sec:obs_spectrum}.
    \item A \textbf{high-energy emission component}, dominated by MeV gamma-rays and potentially ultra-high-energy cosmic rays, with a total energy fraction $\epsilon_{\rm HE} \sim 10^{-4}-10^{-3}$ of the PBH rest-mass energy \cite{Barrau:2014fcg}. 
\end{itemize}
The characteristic burst timescale is set by the light-crossing time of the Schwarzschild radius of the original PBH, $t_{\rm burst}\sim R_S/c = 2GM/c^3\approx 5\times10^{-13}~\text{s}$ for $M\sim10^{23}$~kg. This is far shorter than the millisecond duration of observed FRBs, suggesting that any FRB-like signature would require the burst to be broadened by propagation through the interstellar and intergalactic medium, or the emission timescale to be set by a different physical scale in the outflow. The energy release can reach $E\sim10^{46}$--$10^{47}$~erg for $M\sim10^{23}$~kg. The radio channel could produce FRB-like signatures detectable at cosmological distances if even a small fraction of the available energy is emitted coherently. The gamma-ray and particle channels would be detectable only within $\mathcal{O}(100~\mathrm{Mpc})$ and $\mathcal{O}(50~\mathrm{kpc})$, respectively, with current instruments (see Sec.~\ref{sec:detectability}). The resulting burst rate, therefore, depends sensitively on the PBH mass distribution and on the tunneling parameter $\alpha$, motivating the detailed rate calculations presented in Sec.~\ref{sec:rate}.

\subsection{Implications for Primordial Black Holes}

PBHs formed from early-Universe density fluctuations \cite{Carr:1974nx,Carr:2020gox} can span a wide range of masses. If they comprise a significant fraction of dark matter and if the tunneling process occurs with the lifetime given in Eq.~\eqref{eq:tau}, then PBHs in the above mass range would produce explosive events in the current epoch. The volumetric rate of such bursts is therefore directly tied to the PBH mass function and abundance. As we will show in Sec.~\ref{sec:rate}, the present-day event rate is maximized for masses where $\tau(M)\approx t_0$, and strongly suppressed for both lighter PBHs (already depleted) and heavier PBHs (still metastable).
This sets the stage for linking predictions of loop quantum gravity with astrophysical transient phenomena and provides a testable bridge between quantum gravity, cosmology, and observations of fast radio bursts and other high-energy transients.

\section{Illustrative Detectability Estimates for Planck-Star Bursts}
\label{sec:detectability}

The quantum tunneling of primordial black holes (PBHs) into white holes is predicted to release energy through multiple channels, most prominently a coherent radio flash and a high-energy (gamma-ray or cosmic-ray) component \cite{Barrau:2014yka,Barrau:2014fcg}. In this section, we estimate the observable fluxes for each channel and discuss the corresponding detection horizons with current instrumentation.  Because the emission efficiencies are not calculable from first principles at present, the estimates below should be understood as order-of-magnitude scaling relations that illustrate the parametric dependence on mass and efficiency; they are not firm observational predictions.

\subsection{Emission Energetics}

\subsubsection{Emission Efficiencies in High-Energy and Radio Bands}
\label{sec:efficiencies}

In modeling the observational signatures of black-to-white hole transitions, particularly in the context of Planck-scale remnants or bouncing black holes, we introduce two phenomenological parameters: the efficiency factors $\epsilon_{\rm HE}$ and $\epsilon_{\rm radio}$. These represent the fraction of the total available energy released in a transition that is converted into high-energy (gamma-rays or cosmic rays) and coherent low-frequency (radio) emission, respectively.

\paragraph{High-Energy Emission Efficiency} We adopt a conservative fiducial value $\epsilon_{\rm HE} \sim 10^{-4}$ for the high-energy emission efficiency. This estimate is inspired by analogies with other astrophysical explosive phenomena, such as core-collapse supernovae, gamma-ray bursts (GRBs), and magnetar flares, where only a small fraction of the total energy budget couples to escaping high-energy particles. For example, in typical GRBs, a fraction $\sim 10^{-3} - 10^{-2}$ of the total kinetic energy is converted into prompt gamma-ray emission~\cite{Meszaros:2006rc, Kumar:2014upa}.

In the case of a Planck-star explosion, assuming an available energy of order $E \sim M c^2$, this corresponds to an energy output in gamma-rays of order:
\begin{equation}
    E_{\rm HE} \sim \epsilon_{\rm HE} M c^2 \sim 10^{25} {\rm erg} \left(\frac{\epsilon_{\rm HE}}{10^{-4}} \right)\left(\frac{M}{10^{15}\,\mathrm{g}}\right),
\end{equation}
which, depending on the distance and spectrum, can yield detectable photon fluxes at Earth for a range of masses.

\paragraph{Radio Emission Efficiency} For low-frequency radio emission, we adopt $\epsilon_{\rm radio} \sim 10^{-7}$. This estimate reflects the expected inefficiency of coherent radio emission processes in highly compact, high-curvature environments. Similar values have been invoked in models that attempt to explain FRBs via cataclysmic compact object activity~\cite{Barrau:2014yka, Rees:1977}. In such scenarios, coherent curvature radiation or synchrotron maser emission—possibly arising in a plasma outflow or magnetized shell—can produce a detectable radio burst even with small energy budgets.

The radiated energy in the radio band is therefore approximately:
\begin{equation}
    E_{\rm radio} \sim \epsilon_{\rm radio} M c^2 \sim 10^{22}\  {\rm erg} \left(\frac{\epsilon_{\rm radio}}{10^{-7}} \right)\left(\frac{M}{10^{15}\,\mathrm{g}}\right),
\end{equation}
This suffices to produce observable FRB-like signatures within cosmological volumes for sufficiently numerous or energetic bursts~\cite{Barrau:2014fcg, Barcelo:2017lnx}.

\paragraph{Interpretation and Limitations} These efficiency values should be interpreted as phenomenological placeholders pending a first-principles treatment of the black-to-white hole transition dynamics and radiative processes. They encode our ignorance of the microphysics governing the coupling of the released energy to observable sectors (photons, electrons, etc.), the geometry and plasma environment of the outflow, and the spectrum of emitted quanta. More precise modeling would require assumptions about the quantum gravity resolution of the singularity, the internal structure of the Planck star, and the formation of jets or shells during the bounce. Nonetheless, these choices are consistent with observational benchmarks and allow us to construct illustrative detectability scaling relations.  In practice, our conclusions can be straightforwardly rescaled if future theoretical developments suggest different values for $\epsilon_{\rm HE}$ or $\epsilon_{\rm radio}$.

\begin{figure}[t]
    \centering
    \includegraphics[width=0.75\textwidth]{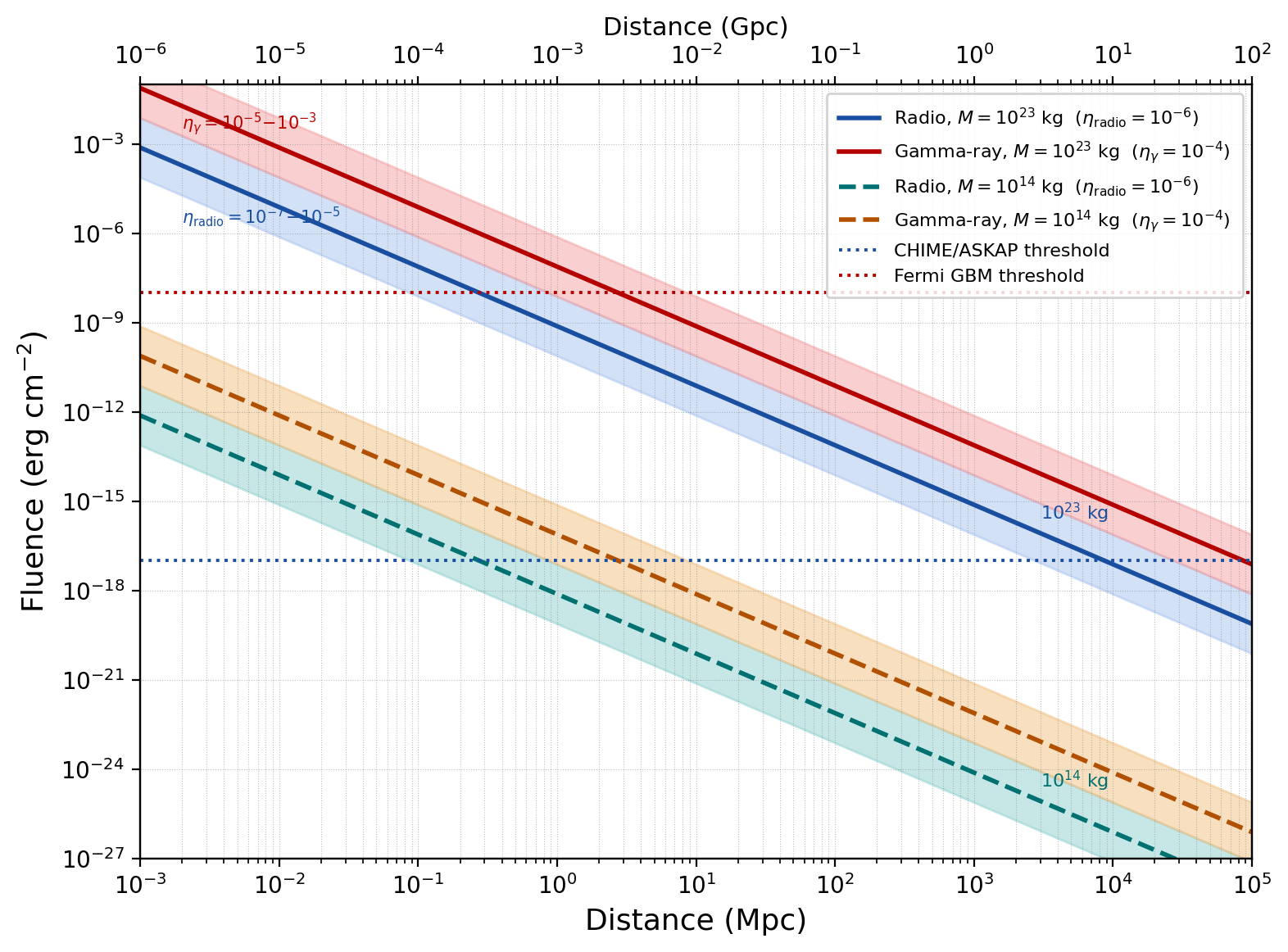}
    \caption{Fluence of Planck-star burst emission in the radio (blue/light blue) and gamma-ray (red/orange)
    channels as a function of distance, for two representative PBH masses:
    $M = 10^{23}$~kg (solid lines) and $M = 10^{14}$~kg (dashed lines, teal and orange).
    Shaded bands indicate the uncertainty in the predicted fluence arising from the
    range of emission efficiencies
    $\epsilon_{\rm radio} \in [10^{-7},\,10^{-5}]$ and
    $\epsilon_{\gamma} \in [10^{-5},\,10^{-3}]$,
    with fiducial values $\epsilon_{\rm radio} = 10^{-6}$ and $\epsilon_{\gamma} = 10^{-4}$
    shown as the central lines.
    Dotted horizontal lines mark the approximate sensitivity thresholds of
    CHIME/ASKAP ($\mathcal{F}_{\rm min} \simeq 10^{-17}$~erg~cm$^{-2}$, radio) and
    Fermi~GBM ($\mathcal{F}_{\rm min} \simeq 10^{-8}$~erg~cm$^{-2}$, gamma-ray).
    The upper axis gives the distance in Gpc.
    For $M = 10^{23}$~kg the radio fluence remains above the CHIME/ASKAP threshold
    out to $\sim 1$~Gpc, while the gamma-ray counterpart becomes undetectable beyond
    $\sim 100$--$300$~Mpc.
    For $M = 10^{14}$~kg both channels fall below current instrument thresholds at
    all but sub-kiloparsec distances, illustrating the steep mass dependence of
    detectability ($\mathcal{F} \propto \epsilon M c^{2} / (4\pi D^{2}$)).}
    \label{fig:detectability}
\end{figure}

\subsection{Fluence and Detection Thresholds}

For a source at luminosity distance $D$, the observed fluence is
\begin{equation}
    F = \frac{E}{4\pi D^2},
\end{equation}
where $E$ is the energy released in the relevant channel. Detection thresholds for representative instruments are:
\begin{itemize}
    \item \textbf{Radio:} $F_{\rm min,radio}\sim10^{-17}$~erg~cm$^{-2}$ for CHIME-like FRB detectors, and $\sim10^{-14}$~erg~cm$^{-2}$ for ASKAP \cite{CHIMEFRB:2021srp}, integrating over $\sim$1~ms at GHz frequencies.
    \item \textbf{Gamma-rays:} $F_{\rm min,\gamma}\sim10^{-8}$~erg~cm$^{-2}$ for short-duration keV--MeV transients detectable by Fermi-GBM or similar instruments \cite{Meegan2009}.
\end{itemize}

The corresponding maximum detection distances are:
\begin{align}
    D_{\rm max,radio} &\approx \sqrt{\frac{E_{\rm radio}}{4\pi F_{\rm min,radio}}}\approx 1~\text{Gpc}\,,\\
    D_{\rm max,\gamma} &\approx \sqrt{\frac{E_{\gamma}}{4\pi F_{\rm min,\gamma}}}\approx 100-300~\text{Mpc}\,.
\end{align}
Thus, coherent radio bursts from Planck-star transitions could be visible out to cosmological distances, comparable to observed FRBs, even with very low radiative efficiency. Gamma-ray counterparts would only be detectable from relatively nearby extragalactic sources, and no coincident emission would be expected from most cosmological events. We note that these detection horizons scale as $D_{\rm max}\propto\epsilon^{1/2}$, so an order-of-magnitude uncertainty in the efficiency translates into a factor $\sim3$ uncertainty in the detection distance. This does not qualitatively change our conclusions, though it does affect rate estimates by up to an order of magnitude.

\subsection{Cosmic Ray Detection Prospects}

Cosmic-ray detection of PSBs is limited to Galactic events ($D\lesssim 50$~kpc), since charged particles undergo diffusion and deflection in extragalactic magnetic fields, washing out any transient signal.  Given the low expected rate of Galactic tunneling events (see Sec.~\ref{sec:rate}), this channel is observationally irrelevant for the foreseeable future.

\subsection{Summary of Detectability}

PSBs are multi-channel phenomena with highly anisotropic observational reach:
\begin{itemize}
    \item The radio emission can be detected to cosmological distances, potentially producing a population of FRB-like events consistent with current observations provided PBHs constitute a non-negligible fraction of the dark matter density.
    \item High-energy gamma-ray counterparts are only observable from within $\sim100$~Mpc, making coincident detection with FRBs unlikely except for nearby events.
    \item Cosmic-ray detection is limited to Galactic events, expected to be very rare.
\end{itemize}
Figure~\ref{fig:detectability} illustrates the expected fluence in the radio and gamma channels as a function of distance, compared to the sensitivity of current instruments. This highlights the unique opportunity to test quantum-gravity inspired PBH tunneling scenarios through the study of extragalactic fast transients.

\section{Fast Radio Bursts: Observational Status and Constraints}
\label{sec:FRB}
\subsection{Global FRB volumetric rate}
\label{sec:obs_rate}

Any viable Planck-star burst (PSB) interpretation of FRBs must reproduce the observed cosmological event rate above survey fluence thresholds.  Recent large-scale surveys, in particular CHIME/FRB operating in the 400--800\,MHz band, imply a volumetric rate
\begin{equation}
R_{\rm FRB} \sim 10^{4} - 10^{5} \ {\rm Gpc^{-3}\ yr^{-1}},
\label{eq:frb_rate_obs}
\end{equation}
for bursts above typical fluence thresholds of order a few Jy\,ms \cite{CHIME_rate_2019,James_2022_rate,Hashimoto_2023_rate}.  The precise normalization depends on bandpass, completeness corrections, and assumptions about the burst energy function, but Eq.~\eqref{eq:frb_rate_obs} provides a robust order-of-magnitude benchmark.

In the white-hole scenario considered here, only a subset of FRBs can plausibly correspond to cataclysmic black-to-white-hole transitions.  We therefore introduce a phenomenological fraction
\begin{equation}
R_{\rm WH} = f_{\rm WH}\, R_{\rm FRB},
\label{eq:fwh_def}
\end{equation}
where $f_{\rm WH}$ denotes the fraction of FRBs attributable to white-hole conversion events.  Because such transitions are intrinsically one-off in the simplest tunneling picture, they cannot account for repeating sources.  Consequently, $f_{\rm WH}$ is bounded from above by the intrinsic non-repeater fraction, $f_{\rm non\text{-}rep}$.

Population analyses indicate that the intrinsic repeater fraction may be large once observational biases are accounted for \cite{CHIME_repeaters_2021,Oppermann_2024_repeaters}, so $f_{\rm non\text{-}rep}$ may be significantly smaller than the fraction of events currently classified as one-off.  A conservative working prior is therefore
\begin{equation}
f_{\rm WH} \lesssim 0.01 - 0.1,
\label{eq:fwh_prior}
\end{equation}
with values approaching $f_{\rm WH}\sim 0.3$ requiring a substantial intrinsically non-repeating population that remains consistent with repetition statistics and follow-up constraints.

In the remainder of this section, we assess whether a white-hole population with rate normalization given by Eq.~\eqref{eq:fwh_def} can simultaneously satisfy temporal, spectral, multiwavelength, and host-galaxy constraints.

\subsection{Temporal properties and repetition}
\label{sec:obs_temporal}

The temporal characteristics of FRBs provide one of the most powerful diagnostics for distinguishing cataclysmic from non-cataclysmic progenitor models.

\subsubsection*{Burst morphology}

Observed bursts typically exhibit millisecond durations and diverse time--frequency morphologies, including multiple components, downward frequency drifting, and microsecond-scale sub-structure \cite{CHIME_morphology_2021,Nimmo_2021_micro}.  Baseband analyses further show that a non-negligible fraction of apparently simple bursts resolve into multiple components at microsecond resolution \cite{CHIME_baseband_2023}, so morphological simplicity at survey resolution does not guarantee an intrinsically single-pulse origin.

In the tunneling scenario, the black-to-white-hole transition is associated with a single explosive event per object, which would yield no repetition, no sustained central-engine activity, and potentially simpler intrinsic time structure.  Consequently, only bursts that are both non-repeating and morphologically simple at high time resolution remain viable white-hole candidates; improved temporal resolution may further reduce this subset.

\subsubsection*{Repetition statistics}

A substantial and growing fraction of localized FRBs are observed to repeat \cite{CHIME_repeaters_2021}, and some repeaters exhibit activity windows or quasi-periodic behavior, supporting the presence of long-lived central engines.  Since white-hole transitions are intrinsically one-off events, repeating sources are incompatible with this channel, and the white-hole contribution to the total FRB rate must satisfy
\begin{equation}
f_{\rm WH} \leq f_{\rm non-rep},
\end{equation}
where $f_{\rm non-rep}$ denotes the fraction of intrinsically non-repeating bursts.  As noted in Sec.~\ref{sec:obs_rate}, sensitivity biases imply that $f_{\rm non-rep}$ is likely smaller than the currently observed one-off fraction \cite{Oppermann_2024_repeaters}, reinforcing $f_{\rm WH}\lesssim\mathcal{O}(0.01\text{--}0.1)$.

Future deep monitoring of apparently non-repeating FRBs will directly tighten constraints on $f_{\rm WH}$ by shrinking the allowed non-repeating subset, and increasingly stringent follow-up campaigns will further test whether any residual one-off population is consistent with the PSB scenario.

\subsection{Radio spectral constraints and implied mass scale}
\label{sec:obs_spectrum}

The radio spectral properties of FRBs probe the characteristic emission scale and therefore, in the white-hole scenario, the underlying black-hole mass.  FRBs are detected from $\sim 100$\,MHz to several GHz \cite{PastorMarazuela_2021_LOFAR,Gajjar_2018_highfreq}, often exhibiting band-limited emission, complex time--frequency structure, and downward-drifting sub-bursts \cite{CHIME_morphology_2021}.  No universal spectral peak is observed across the population.

\subsubsection*{Mass--frequency relation: a benchmark mapping}

In the PSB tunneling framework, a simple dimensional estimate relates the characteristic emission wavelength to the Schwarzschild radius,
\begin{equation}
\lambda_{\rm pk} \sim \kappa R_s = \kappa \frac{2GM}{c^2},
\label{eq:lambda_rs}
\end{equation}
where $\kappa$ is an order-unity model-dependent factor.  The corresponding peak frequency is
\begin{equation}
\nu_{\rm pk} \sim \frac{c}{\lambda_{\rm pk}} = \frac{c^3}{2 G \kappa M},
\label{eq:nu_mass_relation}
\end{equation}
which, upon inversion, gives
\begin{equation}
M \simeq \frac{c^3}{2 G \kappa \nu_{\rm pk}}
\simeq 2.0 \times 10^{29}\,{\rm g}\, 
\left( \frac{1\ {\rm GHz}}{\nu_{\rm pk}} \right)\, \kappa^{-1}.
\label{eq:mass_freq_numeric}
\end{equation}
For the observed FRB frequency interval $\nu_{\rm pk} \sim 0.1\text{--}8$\,GHz, one obtains
\begin{equation}
M \sim (2.5\times10^{28}\text{--}2\times10^{30})\,\kappa^{-1}\ {\rm g},
\label{eq:mass_range_frb}
\end{equation}
i.e.\ roughly Earth-to-Jupiter mass scales for $\kappa\sim 1$.  We emphasize that this mapping assumes prompt, horizon-scale emission and should be regarded as an illustrative benchmark rather than a firm constraint.  If the observed radio signal is instead dominated by reprocessed or coherent plasma emission in the white-hole outflow, the characteristic frequency could be substantially lower than $c/R_s$, decoupling the observed FRB band from the PBH mass.  The rate analysis of Sec.~\ref{sec:rate}---which depends on $M$ and $\alpha$ but not on the emission frequency---is unaffected by this uncertainty; only the spectral identification of specific FRBs with specific PBH masses is model-dependent.

\subsubsection*{Implications for the PBH mass function}

Equation~\eqref{eq:mass_range_frb} implies that FRB-producing PSBs must originate from primordial black holes (PBHs) in a relatively narrow mass window around $10^{29}$\,g (modulo $\kappa$).  A monochromatic PBH mass function would predict clustering of burst peak frequencies around a preferred value; the observed broad distribution of spectral centroids disfavors such a scenario unless $\kappa$ varies substantially from event to event.  A broad PBH mass function could accommodate the frequency diversity but would generically predict a corresponding distribution of intrinsic peak frequencies; detailed modeling would be needed to determine whether propagation effects can sufficiently wash out such correlations.

\subsubsection*{Comparison with astrophysical models}

Astrophysical progenitor models, such as magnetar-based scenarios, attribute spectral structure to plasma emission processes (e.g.\ synchrotron maser or magnetospheric coherent emission).  These mechanisms naturally produce band-limited spectra, frequency drift, and burst-to-burst variability without invoking a single geometric emission scale tied to $R_s$.  While the mass--frequency mapping in Eq.~\eqref{eq:nu_mass_relation} provides an internally consistent interpretation of GHz emission in the PSB framework, current spectral data do not display a universal frequency scale that would uniquely support this hypothesis.

Future statistical tests correlating spectral peak frequency with redshift, burst energetics, or host-galaxy environment could further discriminate between a geometric ($R_s$-driven) origin and plasma-driven emission models.  In the following section we therefore examine whether a PBH population capable of producing bursts in the mass range of Eq.~\eqref{eq:mass_range_frb} can simultaneously reproduce the required volumetric event rate.


\section{Expected Volumetric Rates of PBH Quantum Tunneling Bursts}
\label{sec:rate}


We first consider a monochromatic (delta-function) PBH mass function centred on mass $M$. In the white-hole FRB interpretation, the PBH tunneling rate $R_{\rm PBH}$ is identified with the white-hole contribution $R_{\rm WH}$ defined in Eq.~\eqref{eq:fwh_def}; comparing $R_{\rm PBH}$ with $R_{\rm FRB}$ therefore constrains $f_{\rm WH}$ directly.

The present-day ($z=0$) comoving volumetric rate of quantum tunneling bursts is
\begin{equation}
    R_{\rm PBH}(M,f_{\rm PBH},\alpha)=
    \frac{f_{\rm PBH}\,\rho_{\rm DM,0}}{M}\,
    \frac{1}{\tau_{\rm eff}(M)}\,
    \exp\!\left(-\frac{t_0}{\tau_{\rm eff}(M)}\right),
    \label{eq:rate_density}
\end{equation}
where the exponential factor accounts for depletion over the age of the Universe,
$t_0 \simeq 4.3\times10^{17}\,$s.  We treat Hawking evaporation and white-hole tunneling as independent competing channels, so that the effective lifetime reads:
\begin{equation}
   \frac{1}{\tau_{\rm eff}(M)} =
   \frac{1}{\tau_{\rm evap}(M)} +
   \frac{1}{\tau_{\rm WH}(M)}
   =
   \frac{1}{\kappa_{\!H}}\left(\frac{M_{\rm Pl}}{M}\right)^3\frac{1}{t_{\rm Pl}}
   +
   \frac{1}{\alpha}\left(\frac{M_{\rm Pl}}{M}\right)^2\frac{1}{t_{\rm Pl}},
   \label{eq:tau_eff}
\end{equation}
with $\kappa_{\!H}=5120\pi\simeq1.61\times10^4$ the standard Hawking prefactor\footnote{Not to be confused with the order-unity geometric factor $\kappa$ appearing in the mass--frequency relation of Eq.~\eqref{eq:lambda_rs}.} and
$\rho_{\rm DM,0}\simeq 3.6\times10^{11}\,M_\odot\,\mathrm{Mpc}^{-3}$.
In the tunneling-dominated regime $\tau_{\rm evap}\gg\tau_{\rm WH}$ one has
$\tau_{\rm eff}\simeq\tau_{\rm WH}$, while in the opposite limit
evaporation removes the PBH before tunneling can occur and
 $R_{\rm PBH}$ is strongly suppressed.

\begin{figure}[t]
    \centering
    \includegraphics[width=0.85\linewidth]{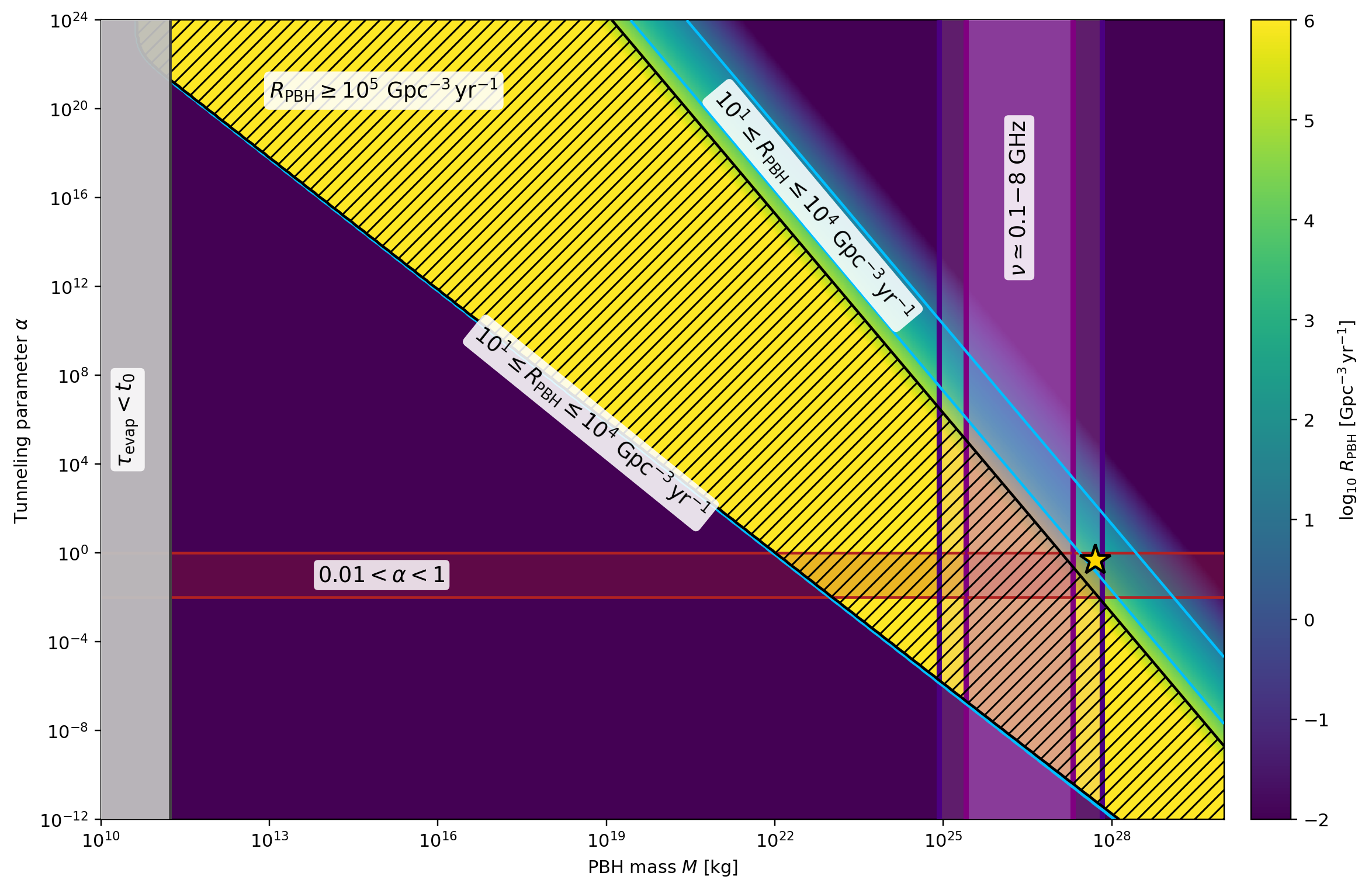}
\caption{
Present-day volumetric burst rate $\log_{10}R_{\rm PBH}$
    [Gpc$^{-3}$\,yr$^{-1}$] in the $(M,\alpha)$ plane, evaluated at
    $z=0$ with $f_{\rm PBH}=1$.  The color scale runs from $-2$
    (dark purple) to $+6$ (bright yellow).  The \emph{diagonal ridge}
    of maximal rate traces the locus $\tau_{\rm eff}(M)\simeq t_0$
    (Eq.~\eqref{eq:ridge}); to its lower-left the population is
    exponentially depleted ($\tau_{\rm eff}\ll t_0$), while to its
    upper-right the instantaneous decay probability is tiny
    ($\tau_{\rm eff}\gg t_0$).
    The \textbf{solid green band} marks the FRB-compatible range
    $10^1\le R_{\rm PBH}\le 10^4\,\mathrm{Gpc^{-3}\,yr^{-1}}$.
    The \textbf{upper-left hatched region} is excluded because the
    predicted burst rate would \emph{exceed} the observed FRB
    volumetric rate ($R_{\rm PBH}>10^5\,\mathrm{Gpc^{-3}\,yr^{-1}}$),
    violating the observational upper bound.
    The \textbf{horizontal red band} indicates the canonical
    Planck-star range $10^{-2}\lesssim\alpha\lesssim 1$; within this
    strip the rate is exponentially suppressed for all masses
    compatible with evaporation constraints.
    The \textbf{vertical blue band} marks the PBH mass window
    $M\sim 10^{24}$--$10^{26}$\,kg for which the Schwarzschild radius
    sets the emission wavelength in the GHz band,
    $\nu_{\rm radio}\sim c/R_S$ (Section~\ref{sec:obs_spectrum}).
    The \textbf{left-boundary hatched strip} marks masses already
    fully depleted by Hawking evaporation ($\tau_{\rm evap}<t_0$).
    The \textbf{star} marks the unique point where three independent
    requirements are simultaneously satisfied: the tunneling parameter
    lies in the canonical Planck-star range
    ($10^{-2}\lesssim\alpha\lesssim 1$, red band), the PBH mass is
    consistent with GHz emission from the Schwarzschild scale
    (vertical blue band, up to an order-unity factor $\kappa$), and
    the predicted burst rate falls within the observed FRB volumetric
    rate (green band).  The coincidence of these three conditions at a
    single point in parameter space is a non-trivial consistency check
    of the Planck-star--FRB hypothesis, though as discussed in the
    text the rate is in practice far below FRB levels once depletion
    and abundance constraints are applied.
    All contours assume a monochromatic mass function; extended
    distributions are discussed in Section~\ref{sec:rate_massfunction}.
}
    \label{fig:RateContour}
\end{figure}

\begin{figure}[t]
    \centering
    \includegraphics[width=0.85\linewidth]{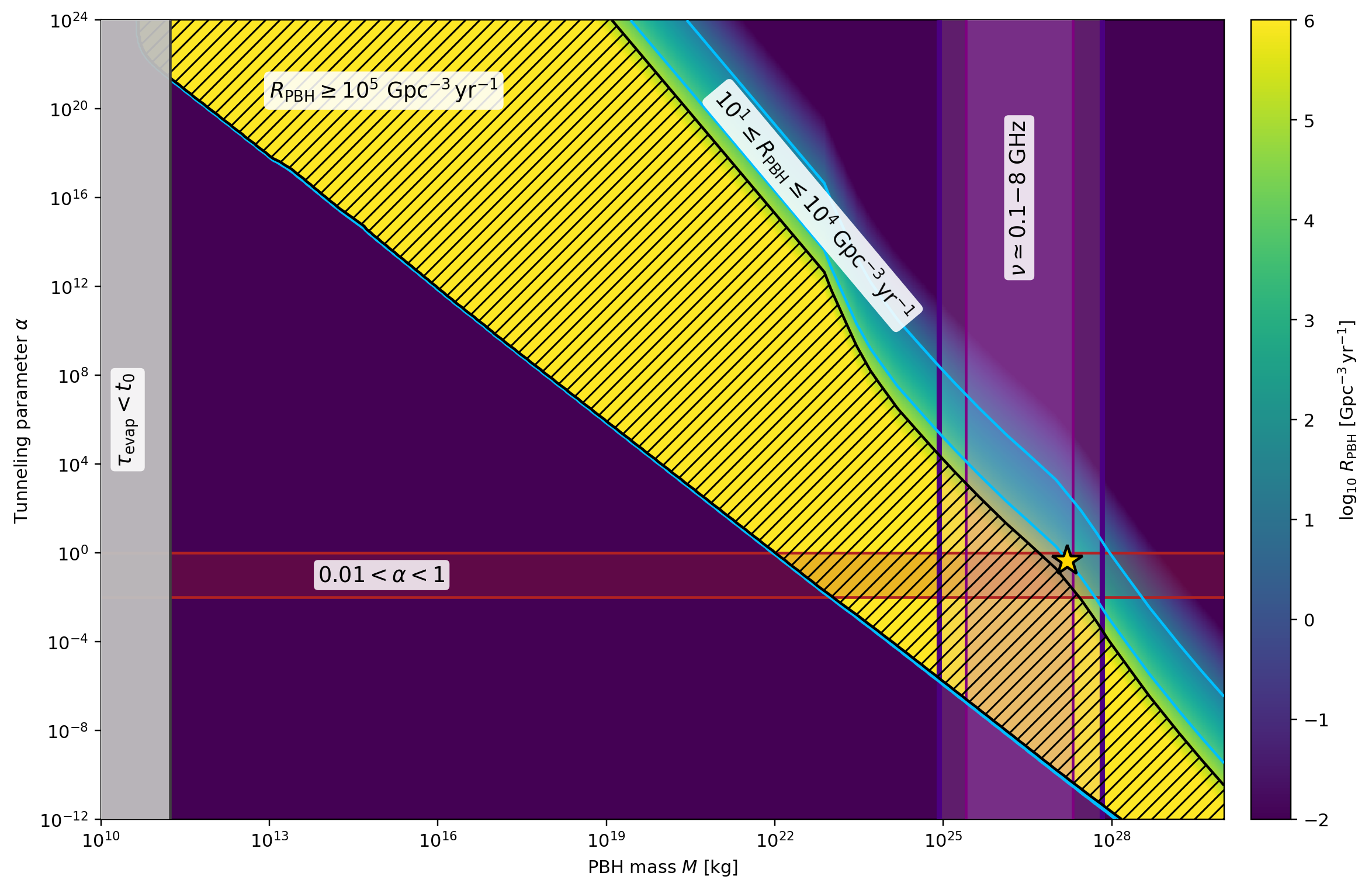}
\caption{%
    Same as Fig.~\ref{fig:RateContour} but with the PBH dark-matter
    fraction set to its mass-dependent observational upper bound
    $f_{\rm PBH,max}(M)$ (inset, grey curve), obtained by
    interpolating current microlensing, CMB spectral-distortion, and
    Hawking-evaporation constraints.  The color scale and all overlays
    are identical to Fig.~\ref{fig:RateContour}, including the hatched
    upper-left region where $R_{\rm PBH}$ would exceed the observed
    FRB rate.
    Abundance constraints enforce $f_{\rm PBH,max}\ll 1$ at
    $M\lesssim 10^{17}$--$10^{20}$\,kg (down to $\sim 10^{-8}$ in
    the most constrained windows), shifting the entire color scale
    downward by the same factor and compressing the diagonal ridge to
    far lower rates.  The FRB-compatible band consequently retreats
    to a narrow wedge confined to very low masses
    ($M\lesssim 10^{11\text{--}12}$\,kg) and very large tunneling
    parameters ($\alpha\gtrsim 10^{18}$), while the rate-excess
    exclusion zone shrinks in proportion.  The \textbf{star} marks
    the same triple-coincidence point as in Fig.~\ref{fig:RateContour};
    once $f_{\rm PBH,max}(M)$ is applied, the predicted rate at that
    point falls well below the FRB-compatible band, illustrating how
    abundance constraints break the apparent consistency of the
    canonical scenario.  Once realistic abundance limits are imposed,
    \emph{no broad region} of the $(M,\alpha)$ plane produces
    cosmologically significant burst rates; any FRB-level scenario
    requires PBHs to saturate their abundance bound precisely in the
    narrow mass window near $10^{11}$--$10^{12}$\,kg.}
    \label{fig:RateContour2}
\end{figure}

\paragraph{Structure of the rate plane.}
Equation~\eqref{eq:rate_density} shows that $R_{\rm PBH}$ is governed
by three competing factors: (i)~the PBH comoving number density
$n_{\rm PBH}\propto f_{\rm PBH}/M$, (ii)~the instantaneous decay
probability $1/\tau_{\rm eff}$, and (iii)~the depletion factor
$\exp(-t_0/\tau_{\rm eff})$.  The rate is maximized when
$\tau_{\rm eff}(M)\simeq t_0$: for $\tau_{\rm eff}\ll t_0$ the
population is exponentially exhausted, while for $\tau_{\rm eff}\gg
t_0$ the per-unit-time decay probability is negligible.  This
condition traces the dominant ridge visible in both
Figs.~\ref{fig:RateContour} and~\ref{fig:RateContour2}, and underlies
all three regimes described below.

\begin{itemize}

\item \textbf{Evaporation-dominated region (low $M$, low $\alpha$).}
For sufficiently small masses the Hawking term in
Eq.~\eqref{eq:tau_eff} dominates, $\tau_{\rm evap}\ll\tau_{\rm WH}$,
giving $\tau_{\rm eff}\simeq\tau_{\rm evap}\propto M^3$.  Since
$\tau_{\rm eff}\ll t_0$, the survival factor
$\exp(-t_0/\tau_{\rm eff})$ is exponentially small, producing the
sharp dark cutoff in the lower-left of both panels.

\item \textbf{Tunneling-dominated region (intermediate $M$, large
$\alpha$).}  When $\tau_{\rm WH}\ll\tau_{\rm evap}$ and
$\tau_{\rm eff}\gtrsim t_0$, the rate scales as
\begin{equation}
  R_{\rm PBH}\propto\frac{f_{\rm PBH}}{\alpha M^3}
  \qquad(\tau_{\rm WH}\ll\tau_{\rm evap},\;\tau_{\rm eff}\gtrsim t_0),
  \label{eq:rate_scaling}
\end{equation}
so iso-rate contours follow $\alpha\propto M^{-3}$, producing the
prominent diagonal structure in both panels.  The ridge of maximal
rate satisfies $\tau_{\rm eff}(M)\simeq t_0$, which in the
tunneling-dominated limit gives
\begin{equation}
  \alpha\simeq\frac{t_0}{t_{\rm Pl}}
  \left(\frac{M_{\rm Pl}}{M}\right)^2,
  \label{eq:ridge}
\end{equation}
up to small corrections from the Hawking term.

\item \textbf{Long-lived regime (high $M$, very large $\alpha$).}
For $\tau_{\rm eff}\gg t_0$, depletion is negligible but
$1/\tau_{\rm eff}$ is tiny, suppressing the rate and producing the
gradual color fading toward the upper-right corner of both panels.

\end{itemize}

In Fig.~\ref{fig:RateContour} ($f_{\rm PBH}=1$), most of the
$(M,\alpha)$ plane is excluded because it produces the wrong rate
in one of two directions.  To the upper left, the predicted rate
\emph{exceeds} the observed FRB volumetric rate
($R_{\rm PBH}>10^5\,\mathrm{Gpc^{-3}\,yr^{-1}}$, hatched region),
ruling out that region by overproduction.
To the lower left, exponential depletion ($\tau_{\rm eff}\ll t_0$)
drives the rate to negligible values.  To the upper right, the
long effective lifetime ($\tau_{\rm eff}\gg t_0$) renders the
instantaneous decay probability tiny.
Only the narrow FRB-compatible band (solid green) threads between
these three failure modes.
Within the canonical Planck-star strip
$10^{-2}\lesssim\alpha\lesssim 1$ (horizontal red band), the
effective lifetime falls far below $t_0$ for every mass that
survives evaporation, so the present-day rate is exponentially
suppressed throughout.
Achieving FRB-compatible rates while remaining below the
overproduction bound requires
$\alpha\gtrsim 10^{18}$ and $M\lesssim 10^{11\text{--}12}$\,kg.

The star in Fig.~\ref{fig:RateContour} marks the unique point where
the canonical Planck-star range ($10^{-2}\lesssim\alpha\lesssim 1$),
the Schwarzschild-scale GHz emission band
(Section~\ref{sec:obs_spectrum}), and the FRB-compatible rate band
all intersect.  This apparent consistency, however, is an artifact
of the optimistic $f_{\rm PBH}=1$ assumption: once realistic abundance
constraints are applied (Fig.~\ref{fig:RateContour2}), the rate at
that point drops well below FRB levels.

\paragraph{Including mass-dependent $f_{\rm PBH}$ bounds.}
The assumption $f_{\rm PBH}=1$ is highly optimistic.  In
Fig.~\ref{fig:RateContour2} we instead normalize the rate to
$f_{\rm PBH,max}(M)$, where $f_{\rm PBH,max}(M)$ denotes the most stringent upper bound
on the PBH dark-matter fraction at mass $M$
\cite{Carr:2020gox,Green:2020jor}:
\begin{equation}
  R_{\rm PBH}(M,\alpha)=
  \frac{f_{\rm PBH,max}(M)\,\rho_{\rm DM,0}}{M}\,
  \frac{1}{\tau_{\rm eff}(M)}\,
  \exp\!\left(-\frac{t_0}{\tau_{\rm eff}(M)}\right).
  \label{eq:rate_fpbhmax}
\end{equation}
At $M\lesssim 10^{17}$--$10^{20}$\,kg, microlensing, CMB, and
evaporation constraints enforce $f_{\rm PBH,max}\ll 1$---frequently
$10^{-3}$--$10^{-8}$ or smaller---shifting the color scale downward
by the same factor and compressing the diagonal ridge to far lower
values.  The FRB-compatible band retreats to a narrow wedge at very
low masses and extreme $\alpha$, while the rate-excess exclusion zone
shrinks in proportion.  

The essential conclusion from Fig.~\ref{fig:RateContour2} is that
once realistic abundance constraints are imposed, the failure mode
shifts: whereas the benchmark $f_{\rm PBH}=1$ scan
(Fig.~\ref{fig:RateContour}) excludes parameter space in both
directions---overproduction at large rates and underproduction at
small rates---the realistic scan is dominated almost entirely by
underproduction.  Abundance bounds suppress the rate over most of
the plane, collapsing the FRB-compatible region to a narrow wedge at
$M\sim 10^{11}$--$10^{12}$\,kg and $\alpha\gtrsim 10^{18}$,
where abundance bounds are least restrictive and the tunneling
lifetime satisfies $\tau_{\rm eff}\simeq t_0$.  Any FRB-level
scenario therefore requires PBHs to saturate their abundance upper
bounds in precisely this narrow mass window.  The scenario is not
ruled out, but it is tightly confined and requires substantial
fine-tuning in both mass and tunneling parameter.

\paragraph{Redshift-dependent rate density.}

To leading order, the comoving PBH number density is conserved:
\begin{equation}
    n_{\rm PBH}^{\rm com} = \frac{f_{\rm PBH}\rho_{\rm DM,0}}{M},
\end{equation}
independent of redshift. The observed burst rate per unit comoving volume per unit 
\emph{observer time}, however, differs from the source-frame rate for two reasons:

\begin{enumerate}
\item A $(1+z)^{-1}$ time-dilation factor between source and observer clocks.
\item The survival fraction must be evaluated at the cosmic age $t(z)$ rather than $t_0$.
\end{enumerate}

The comoving rate density at redshift $z$ is therefore
\begin{equation}
    R_{\rm PBH}(z) =
    \frac{f_{\rm PBH}\rho_{\rm DM,0}}{M}
    \frac{1}{(1+z)\,\tau_{\rm eff}(M)}
    \exp\!\left(-\frac{t(z)}{\tau_{\rm eff}(M)}\right),
    \label{eq:rate_z_full}
\end{equation}
where $t(z)$ is the cosmic age at redshift $z$.

The survey-averaged volumetric rate out to $z_{\rm max}$ is
\begin{equation}
    \bar{R}_{\rm PBH} =
    \frac{\displaystyle\int_0^{z_{\rm max}} 
    R_{\rm PBH}(z)\,\frac{dV_c}{dz}\,dz}
    {\displaystyle\int_0^{z_{\rm max}} 
    \frac{dV_c}{dz}\,dz},
    \label{eq:rate_avg_full}
\end{equation}
with $dV_c/dz = 4\pi d_c^2(z)\,c/H(z)$.

We now examine how this modifies the structure of the $(M,\alpha)$ rate plane shown in 
Figs.~\ref{fig:RateContour} and~\ref{fig:RateContour2}, noting, first off, that redshift evolution modifies the normalization of the rate but does not
qualitatively alter the structure of the $(M,\alpha)$ plane.

First, the explicit $(1+z)^{-1}$ time-dilation factor reduces the
observer-frame rate from high redshift.
Second, the survival factor becomes
$\exp[-t(z)/\tau_{\rm eff}]$.
For $\tau_{\rm eff}\sim t_0$ (the region of maximal present-day rate),
$t(z)<t_0$ at $z>0$, so the exponential suppression is weaker in the past.
The intrinsic source-frame burst probability was therefore higher at earlier
cosmic times.
However, this enhancement is partially offset by time dilation and by the
finite comoving volume element at large $z$.

For $\tau_{\rm eff}\ll t_0$ (low-mass region),
the PBH population is already exponentially depleted by $z=0$.
At higher redshift the depletion is less severe,
but because $\tau_{\rm eff}$ is extremely short,
most events occurred at very early epochs ($z\gg 1$),
well outside the range probed by FRB surveys.
These early explosions instead contribute to the cumulative diffuse
$\gamma$-ray background.
Thus the low-mass portion of the plane is more strongly constrained
by integrated energy injection than by transient rates.

For $\tau_{\rm eff}\gg t_0$ (high-mass region),
the redshift dependence is negligible:
the decay probability remains small at all $z\lesssim 2$,
and the cosmological averaging in Eq.~\eqref{eq:rate_avg_full}
changes the rate only at the $\mathcal{O}(1)$ level.

Numerically, integrating Eq.~\eqref{eq:rate_avg_full} to
$z_{\rm max}\simeq 1$--2 increases the effective volumetric rate
by at most a factor of a few in the $\tau_{\rm eff}\sim t_0$ band,
since the comoving volume grows rapidly with redshift while the
$(1+z)^{-1}$ factor moderates the contribution from high $z$.
This modest enhancement is utterly insufficient to compensate for the
$\gtrsim 10^{10}$--$10^{12}$ order-of-magnitude deficit between the
canonical $\alpha\sim\mathcal{O}(1)$ predictions and FRB rates.

\subsection{Extended Mass Function}
\label{sec:rate_massfunction}

Physical PBH formation mechanisms generically produce extended mass distributions rather than a monochromatic spectrum. A lognormal mass function is well motivated by inflationary models with a sharp feature in the primordial power spectrum \cite{Carr:2020xqk}, and is widely used as a benchmark:
\begin{equation}
    \frac{dn_{\rm PBH}}{d\ln M} = \frac{f_{\rm PBH}\,\rho_{\rm DM,0}}{\sqrt{2\pi}\,\sigma}
    \exp\!\left(-\frac{\ln^2(M/M_c)}{2\sigma^2}\right),
    \label{eq:lognormal}
\end{equation}
where $M_c$ is the peak mass and $\sigma$ is the width parameter in decades of mass. The monochromatic case is recovered in the limit $\sigma \to 0$.

The total volumetric burst rate for an extended mass function is obtained by integrating the per-mass rate over the distribution:
\begin{equation}
    R_{\rm PBH}^{\rm tot}(f_{\rm PBH},\alpha) = \int_{M_{\rm min}}^{M_{\rm max}}
    \frac{1}{\tau_{\rm eff}(M)}\,e^{-t_0/\tau_{\rm eff}(M)}\,
    \frac{dn_{\rm PBH}}{d\ln M}\,\frac{d\ln M}{M}.
    \label{eq:rate_extended}
\end{equation}

The integrand is sharply peaked around the mass $M_{\rm peak}$ where $\tau_{\rm eff}(M_{\rm peak}) = t_0$. For a lognormal distribution with width $\sigma$, the effective rate relative to the monochromatic case at $M_c$ is enhanced by a factor that depends on how much of the distribution lies near $M_{\rm peak}$.

We can assess this analytically. The rate $R(M) \equiv \tau_{\rm eff}^{-1}(M)\,e^{-t_0/\tau_{\rm eff}(M)}$ is a sharply peaked function of $M$ centred at $M_{\rm peak}$, with a characteristic width in $\ln M$ of order unity (since $\tau_{\rm eff}\propto M^2$ and the peak condition $\tau_{\rm eff}=t_0$ determines a unique mass for each $\alpha$). The convolution with the lognormal therefore gives:
\begin{itemize}
    \item If the lognormal peak $M_c \approx M_{\rm peak}$, the integrated rate is enhanced over the monochromatic estimate by a factor $\sim \exp(\sigma^2/2)$ from the wings of the distribution that overlap the favorable rate region. For $\sigma = 1$ (a factor of $e$ in mass per standard deviation), this is a factor $\sim 1.6$---negligible given the many-orders-of-magnitude shortfall.
    \item If $M_c \neq M_{\rm peak}$, the distribution places most of its weight away from the resonant mass and the integrated rate is suppressed exponentially, behaving as $e^{-\ln^2(M_c/M_{\rm peak})/2\sigma^2}$ relative to having $M_c = M_{\rm peak}$.
    \item For very broad distributions ($\sigma \gtrsim 3$), essentially spanning the whole mass range, the integrated rate approaches a mass-averaged value $\langle R(M) \rangle$ that is parametrically the same order as the monochromatic estimate at $M_{\rm peak}$, since the dominant contribution still comes from the narrow range near $M_{\rm peak}$.
\end{itemize}

The essential conclusion is that a lognormal mass function does not rescue the rate: the rate integrand $R(M)$ is intrinsically a narrow function of mass (its relative width is $\Delta\ln M \sim 1$), so broadening the mass distribution provides at most a factor $\sim \exp(\sigma^2/2)$ enhancement---far less than the $10^{11}$ or more orders of magnitude needed to reach FRB rates for standard $\alpha$. In the FRB-compatible region ($\alpha \gtrsim 10^{18}$, $M \lesssim 10^{11}$~kg), the result is similarly insensitive to the width $\sigma$: what matters is whether the distribution places any weight at the required low masses, which is independently constrained by Hawking evaporation limits. An extended mass function therefore does not qualitatively change the conclusions from the monochromatic analysis. The WH-FRB scenario therefore is not qualitatively affected by moving to a broader mass distribution.

\subsection{Theoretical Status of the Tunneling Parameter}
\label{sec:alpha_theory}

As alluded to above, the timescale for the quantum tunneling of a black hole  to a white hole  state is commonly parametrized as in Eq.~\eqref{eq:intro_tau} \cite{Haggard:2014rza,Rovelli:2014cta,Christodoulou:2016vny},
where $M$ is the PBH mass, $M_{\rm Pl}$ and $t_{\rm Pl}$ are the Planck mass and time, respectively, and $\alpha$ is a dimensionless parameter encoding the detailed quantum gravitational dynamics of the transition.

In heuristic Planck-star models \cite{Haggard:2014rza,Rovelli:2014cta}, $\alpha$ is assumed to be $\mathcal{O}(0.01-1)$, corresponding to a bounce time comparable to the naive dynamical timescale for an interior quantum transition. However, the value of $\alpha$ is not derived from first principles and remains a free parameter subject to large theoretical uncertainties.

In principle, $\alpha$ could be computed from a complete non-perturbative quantum gravity theory by evaluating the transition amplitude:
\begin{equation}
    \Gamma(M) \sim \frac{1}{t_{\rm Pl}}\,e^{-S_{\rm E}(M)},
\end{equation}
where $S_{\rm E}(M)$ is the Euclidean gravitational action for an instanton mediating the BH$\rightarrow$WH transition. In this framework, $\alpha$ encapsulates the ratio between the naive bounce timescale and the exponentially suppressed tunneling probability:
\begin{equation}
    \alpha \propto \frac{e^{S_{\rm E}(M)}}{(M/M_{\rm Pl})^2}.
\end{equation}
The instanton action is expected to scale with the BH entropy, $S\sim M^2/M_{\rm Pl}^2$, but its precise coefficient and dependence on the internal quantum state are unknown.

Alternative approaches include:
\begin{itemize}
    \item \textbf{Loop quantum gravity (LQG):} Discretized quantum geometry leads to a bounce replacing the singularity, with a transition time $\tau\sim(M/M_{\rm Pl})^2 t_{\rm Pl}$ and $\alpha\sim O(1)$ \cite{Christodoulou:2016vny}.
    \item \textbf{Effective path integrals:} Non-perturbative saddle points in the gravitational path integral could, in principle, yield an explicit value for $\alpha$ via $S_{\rm E}(M)$.
    \item \textbf{Microstate overlap:} The matrix element between the BH and WH Hilbert space sectors, $\langle{\rm WH}|T|{\rm BH}\rangle$, could be small if the number of WH microstates is much lower than that of BHs, effectively enhancing $\alpha$.
\end{itemize}

At present, $\alpha$ is an effective parameter absorbing all quantum gravitational uncertainties in the tunneling amplitude. A precise calculation awaits a UV-complete theory of quantum gravity with well-defined BH and WH quantum states and their transition dynamics.

\subsection{Memory-burden scenario: sequential evolution}\label{sec:memory_burden}
 
The baseline analysis of Sec.~\ref{sec:rate} assumes that Hawking
evaporation and quantum tunneling compete simultaneously.  An
alternative possibility---not a prediction of the baseline framework,
but a physically motivated scenario that brackets the theoretical
uncertainty in the evaporation--tunneling interplay---arises if black
holes store quantum information in a large number of internal
microstates.  In such scenarios, Hawking evaporation induces a
dynamical backreaction---\emph{memory burden}---that slows and
eventually halts mass loss once a significant fraction of the initial
mass has been radiated~\cite{Dvali:2018mem,Dvali:2020wft}.
The essential mechanism is that information loaded in the black hole's internal
degrees of freedom resists its decay: the quantum state of the memory modes
backreacts on the evaporating master mode, creating an effective energy barrier
that suppresses further emission~\cite{Dvali:2024hsb}.
Applied to PBHs, the memory-burden effect opens a new mass window for
dark-matter candidacy below the standard Hawking evaporation
limit~\cite{Alexandre:2024nuo,Thoss:2024hsr}, since
black holes that would have fully evaporated in the semiclassical picture
can instead survive to the present epoch as stabilized remnants.
 
This qualitatively changes the interplay between evaporation and tunneling, replacing the competitive picture of Sec.~\ref{sec:rate} with a \emph{sequential} evolution in which tunneling is forbidden until the evaporation phase completes.  We describe here the minimal implementation and its consequences for the burst rate.
 
\subsubsection*{Dynamical framework}
 
In the minimal memory-burden model~\cite{Dvali:2020wft,Alexandre:2024nuo}, evaporation
proceeds in the standard semiclassical regime until the black hole has lost
approximately half of its initial mass, at which point quantum backreaction
becomes unavoidable.  The black hole then stabilizes at a remnant mass
$M_\ast=M/2$ on a timescale
\begin{equation}
\tau_{\rm half}=\frac{7}{8}\,\tau_{\rm evap}(M)
= \frac{7}{8}\,\kappa_{\!H}\left(\frac{M}{M_{\rm Pl}}\right)^{\!3} t_{\rm Pl}\,,
\label{eq:tau_half}
\end{equation}
after which the black hole is stabilized against further Hawking emission.
More precisely, the post-burden evaporation rate is suppressed by a power
of the black hole entropy, $\dot{M}\propto S^{-k}$ with $k\geq 1$ an
integer~\cite{Dvali:2020wft,Dvali:2024hsb}; for the PBH mass scales
of interest here ($S\gg 1$), this suppression is so extreme that the
remnant is effectively stable on cosmological timescales unless another
decay channel---such as white-hole tunneling---becomes available.
 
White-hole tunneling then becomes the dominant decay channel, with lifetime
\begin{equation}
\tau_{\rm WH}(M_\ast)=\alpha\left(\frac{M_\ast}{M_{\rm Pl}}\right)^{\!2} t_{\rm Pl}
= \frac{\alpha}{4}\left(\frac{M}{M_{\rm Pl}}\right)^{\!2} t_{\rm Pl}\,.
\label{eq:tau_wh_remnant}
\end{equation}
The total effective lifetime is the \emph{sum} rather than the harmonic mean:
\begin{equation}
\tau_{\rm eff}^{\rm (seq)}(M) = \tau_{\rm half}(M) + \tau_{\rm WH}(M_\ast)\,.
\label{eq:tau_seq}
\end{equation}
 
\subsubsection*{Critical mass and rate structure}
 
A tunneling burst at $z=0$ requires $\tau_{\rm eff}^{\rm (seq)}\lesssim t_0$.
Since $\tau_{\rm half}\propto M^3$ grows faster than $\tau_{\rm WH}\propto M^2$,
the evaporation phase alone imposes a hard upper mass threshold:
\begin{equation}
\tau_{\rm half}(M) < t_0
\quad\Longrightarrow\quad
M < M_c \equiv \left(\frac{8\,t_0}{7\,\kappa_{\!H}\,t_{\rm Pl}}\right)^{\!1/3} M_{\rm Pl}
\sim 10^{11\text{--}12}\;\mathrm{kg}\,.
\label{eq:Mc_memory}
\end{equation}
Above $M_c$, the evaporation phase has not yet completed and the burst
rate vanishes identically at $z=0$, regardless of $\alpha$.
 
Below $M_c$, the present-day rate takes the same exponential form as
Eq.~\eqref{eq:rate_density} but with $\tau_{\rm eff}$ replaced by
$\tau_{\rm eff}^{\rm (seq)}$:
\begin{equation}
R_{\rm PBH}^{\rm (seq)}(M,\alpha) =
\frac{f_{\rm PBH}\,\rho_{\rm DM,0}}{M}\;
\frac{1}{\tau_{\rm eff}^{\rm (seq)}(M)}\;
\exp\!\left(-\frac{t_0}{\tau_{\rm eff}^{\rm (seq)}(M)}\right).
\label{eq:rate_seq}
\end{equation}
An observable rate requires $\tau_{\rm WH}(M_\ast)\sim t_0 - \tau_{\rm half}(M)$,
confining $\alpha$ to a narrow band for each $M < M_c$.
The broad diagonal ridge of the competitive case
(Figs.~\ref{fig:RateContour}--\ref{fig:RateContour2}) therefore collapses
to a sharply bounded strip near the evaporation boundary.  A further
consequence is that diffuse gamma-ray constraints become largely
irrelevant over most of the $(M,\alpha)$ plane, since the energy injection rate
$R\,\epsilon_\gamma\,Mc^2$ is negligible everywhere except within the small
surviving wedge.
 
We note that the memory-burden framework also modifies PBH abundance
constraints derived from big-bang nucleosynthesis and CMB spectral
distortions, since the extended lifetime changes the epoch and spectrum
of particle injection~\cite{Alexandre:2024nuo,Thoss:2024hsr}.  The
$f_{\rm PBH,max}(M)$ bounds used in our baseline analysis
 are derived under the standard Hawking picture;
in the memory-burden scenario these bounds would be relaxed for
$M\lesssim 10^{14}$--$10^{17}$\,g, potentially widening the viable
parameter space at the lowest masses.  A self-consistent treatment
incorporating memory-burden-modified abundance limits is beyond the
scope of this work but would strengthen the case for a surviving
low-mass wedge.
 
\subsubsection*{Comparison with the competitive case}
 
The key qualitative differences are:
(i)~the viable parameter space shrinks from a broad diagonal ridge to a narrow
strip anchored at $M\lesssim M_c$;
(ii)~early-time tunneling is forbidden, so cumulative energy-injection
constraints are significantly weakened;
and (iii)~the predicted burst population is sharply peaked at a single mass scale
rather than spread along a ridge, producing a more distinctive---and more
falsifiable---observational signature.
Whether the memory-burden picture is realized in nature depends on
unresolved questions in black-hole quantum information
theory~\cite{Dvali:2024hsb}; we include
it here as a physically motivated alternative that brackets the
uncertainty in the evaporation--tunneling interplay.

\subsection{Non-zero spin effects}\label{sec:spin}

PBHs formed from radiation-era density fluctuations carry a small but
nonzero angular momentum, with characteristic dimensionless spin
$\chi_0\lesssim\mathcal{O}(10^{-2})$ at formation~\cite{Chiba:2017rvs}.
Spin reduces the Hawking temperature relative to the Schwarzschild
value by a factor $h(\chi)\approx 1-\chi^2/4+\mathcal{O}(\chi^4)$,
slightly lengthening the evaporation time by a fractional correction
$\delta_\chi\sim\mathcal{O}(\chi_0^2)$; Page~\cite{Page:1976df} showed
numerically that this amounts to $\lesssim 2\%$ for $\chi_0=0.1$ and
$\lesssim 0.02\%$ for $\chi_0=0.01$, since the black hole spins down
to $\chi\approx 0$ after losing only $\lesssim 25\%$ of its mass.
The corresponding shift in the critical mass,
$M_{\rm crit}(\chi_0)\approx M_{\rm crit}(0)(1-\delta_\chi)$, is at
most $\sim 2\%$---entirely negligible relative to the
orders-of-magnitude uncertainties in $\alpha$.  The spin dependence
of the tunneling amplitude itself is unknown, as no computation of the
instanton action $S_{\rm E}(M,\chi)$ for Kerr black holes exists in
the LQG framework, but symmetry arguments suggest the correction is
$\mathcal{O}(\chi_0^2)$.  The Schwarzschild approximation used
throughout this work is therefore well-justified for the relevant PBH
population.

\section{Gamma-ray and multiwavelength constraints}
\label{sec:obs_gamma}

Multiwavelength observations provide a largely independent test of the white-hole tunneling interpretation of FRBs.  In particular, searches for prompt gamma-ray counterparts, constraints from the diffuse gamma-ray background, and prospects for gravitational-wave detection are all relevant, given the potentially explosive nature of a black-to-white-hole transition.

\subsection*{Prompt gamma-ray fluence}

Consider a white-hole transition originating from a black hole of mass $M$.  The total available rest-mass energy is $E_{\rm tot} = Mc^2$.  For a representative mass scale compatible with GHz radio emission under the assumption $\lambda_{\rm pk}\sim R_s$ (cf.\ Sec.~\ref{sec:obs_spectrum}),
\begin{equation}
M \sim 10^{23}\ {\rm g},
\qquad
E_{\rm tot} \simeq 9 \times 10^{43}\ {\rm erg}.
\end{equation}
Let $\epsilon_\gamma$ denote the fraction of this energy emitted in gamma rays within the observational band of interest (corresponding to the high-energy efficiency $\epsilon_{\rm HE}$ of Sec.~\ref{sec:efficiencies}).  The observed fluence at luminosity distance $d_L$ is
\begin{equation}
\mathcal{F}_\gamma \simeq \frac{(1+z)\,\epsilon_\gamma\, M c^2}{4\pi d_L^2}.
\label{eq:gamma_fluence}
\end{equation}

At $z\simeq 0.5$, corresponding roughly to the median redshift of the localized FRB population and to $d_L\simeq 2.8$\,Gpc in standard $\Lambda$CDM cosmology, this yields
\begin{equation}
\mathcal{F}_\gamma \simeq 1.4 \times 10^{-13}\,
\epsilon_\gamma\ {\rm erg\ cm^{-2}}.
\label{eq:gamma_numeric}
\end{equation}
Searches for prompt gamma-ray counterparts to FRBs using instruments such as \textit{Fermi}/GBM and \textit{Swift}/BAT typically achieve sensitivity limits of order $\mathcal{F}_{\rm lim}\sim 10^{-8}$\,erg\,cm$^{-2}$ in the 10--1000\,keV range for short transients \cite{Scholz_2020_GBM,Laha_2022_gamma}.  Even for $\epsilon_\gamma\sim 1$, the predicted fluence lies approximately five orders of magnitude below current thresholds.  Non-detections of prompt gamma-ray counterparts to cosmological FRBs therefore do not meaningfully constrain white-hole transitions at $M\sim 10^{23}$\,g.

\subsection*{Diffuse gamma-ray background}

Although individual events are too faint to detect at cosmological distances, a large population could in principle contribute to the isotropic gamma-ray background (IGRB).  The cumulative energy injection rate per comoving volume is
\begin{equation}
\dot{\rho}_\gamma \sim f_{\rm WH}\, R_{\rm FRB}\, \epsilon_\gamma\, M c^2.
\end{equation}
Using $R_{\rm FRB}\sim 10^4$--$10^5$\,Gpc$^{-3}$\,yr$^{-1}$ and $M\sim 10^{23}$\,g, one finds
\begin{equation}
\dot{\rho}_\gamma \sim 3 \times 10^{38}\,
f_{\rm WH}\, \epsilon_\gamma\ 
{\rm erg\ Gpc^{-3}\ yr^{-1}},
\end{equation}
which is several orders of magnitude below the total gamma-ray emissivity inferred from the IGRB.  Combined with the constraint $f_{\rm WH}\lesssim 0.1$ from Sec.~\ref{sec:FRB}, the diffuse gamma-ray bound is doubly suppressed, and current IGRB measurements do not constrain the parameter range considered here.

\subsection*{Gravitational-wave diagnostics}
\label{sec:obs_gw}

Gravitational waves offer an additional multimessenger probe of the white-hole scenario, through two complementary channels.

\subsubsection*{Prompt gravitational-wave emission from the transition}

A cataclysmic black-to-white-hole transition is expected to produce a prompt burst of gravitational radiation.  On dimensional grounds, the characteristic frequency of this emission is set by the light-crossing time of the Schwarzschild radius,
\begin{equation}
f_{\rm GW}^{\rm burst} \sim \frac{c}{R_s} = \frac{c^3}{2\,G\,M}
\simeq 1\,{\rm GHz}\left(\frac{10^{23}\ {\rm g}}{M}\right),
\label{eq:fgw_burst}
\end{equation}
which, for the mass range $M\sim 10^{23}$\,g implied by the spectral analysis of Sec.~\ref{sec:obs_spectrum}, falls in the GHz band.  (The numerical prefactor differs from Eq.~\eqref{eq:nu_radio} by factors of $\pi$ that are absorbed into the order-unity geometric factor $\kappa$ of Eq.~\eqref{eq:lambda_rs}.)  This lies far above the sensitivity range of current and planned laser-interferometer observatories (LIGO/Virgo/KAGRA at $\sim$10--$10^4$\,Hz; LISA at $\sim$10$^{-4}$--10$^{-1}$\,Hz; Einstein Telescope and Cosmic Explorer at $\sim$1--$10^4$\,Hz), and also above the reach of proposed high-frequency gravitational-wave (HFGW) detectors targeting the kHz--MHz band \cite{Aggarwal_2021_HFGW}.  A direct detection of the prompt GW burst from an individual transition therefore appears well beyond current or near-future experimental capabilities.

\subsubsection*{PBH binary mergers}

If a cosmologically significant population of PBHs in the mass range of Eq.~\eqref{eq:mass_range_frb} ($\sim 10^{-5}$--$10^{-3}\,M_\odot$) exists, a fraction will form binaries and merge \cite{Bird_2016_PBHmerger,Sasaki_2016_PBHmerger}, producing gravitational-wave signals at ultra-high frequencies ($10^{4}$--$10^{6}$\,Hz).  Current experimental limits from bulk acoustic wave resonators~\cite{MAGE_2025} and projected sensitivities of resonant cavity experiments~\cite{Aggarwal_2021_HFGW,Domcke_2022_HFGW} remain orders of magnitude above predicted rates~\cite{Franciolini_2022_HFGW}.  This channel is not a direct probe of the tunneling event itself but could independently confirm or rule out the requisite progenitor population.

\subsection*{Other wavelength diagnostics}

In contrast to magnetar-based models, white-hole transitions are not expected to produce persistent nebulae, long-lived X-ray counterparts, or repeated activity.  Localized FRBs frequently occur in star-forming galaxies and in some cases are associated with persistent radio sources---features more naturally accommodated within astrophysical progenitor scenarios.

A decisive observational signature for a white-hole channel would be the detection of a non-repeating FRB accompanied by a prompt high-energy or gravitational-wave transient inconsistent with magnetar activity, particularly in a host environment tracing dark matter rather than star formation.  No such event has been conclusively identified to date.

\medskip

In summary, current gamma-ray, gravitational-wave, and multiwavelength observations neither strongly constrain nor positively support a white-hole interpretation at the mass scale compatible with GHz emission.  The absence of prompt high-energy counterparts is consistent with the low expected fluence of individual cosmological events.  Prompt gravitational radiation from the transition itself falls in the GHz band, far above the reach of existing detectors, while gravitational waves from PBH binary mergers---though not a direct probe of the tunneling event---offer a complementary pathway to establishing or ruling out the requisite PBH population.  In the near term, repetition statistics, host-galaxy demographics, and increasingly stringent multiwavelength follow-up campaigns provide the most stringent empirical guidance.

\subsection{Host-galaxy demographics and redshift evolution}
\label{sec:obs_hosts}

The redshift evolution and host-galaxy properties of FRBs provide a powerful discriminator between progenitor channels that trace stellar processes and those that follow the cosmological dark matter distribution.  In the white-hole tunneling framework, events are associated with primordial black holes and are therefore expected, to leading order, to trace the large-scale dark matter density rather than recent star formation.

\subsubsection*{Observed host properties}

Localized FRBs have been identified in a wide variety of host galaxies, including low-mass star-forming dwarfs, massive spirals, and, in a smaller number of cases, more passive systems \cite{Bhandari_2020_hosts,Heintz_2020_hosts}.  Several repeating FRBs are associated with actively star-forming regions, and in at least one case with a persistent compact radio source \cite{Chatterjee_2017_persistent}.  Population studies suggest that the volumetric FRB rate evolves with redshift in a manner broadly consistent with, or mildly tracking, the cosmic star formation rate density \cite{Hashimoto_2023_rate,James_2022_rate}; there is no compelling evidence for a purely dark-matter-tracing evolution.

\subsubsection*{Rate evolution: dark matter vs.\ star-formation scaling}

In the white-hole scenario, with negligible delay time relative to cosmological timescales, the comoving event rate scales approximately as
\begin{equation}
R_{\rm WH}(z) \propto \rho_{\rm DM}(z),
\end{equation}
modulo depletion effects associated with the PBH lifetime.  Since the comoving dark matter density is nearly constant, this implies weak intrinsic redshift evolution apart from cosmic time dilation and possible exponential depletion factors.

By contrast, astrophysical progenitors such as magnetars formed in core-collapse supernovae scale approximately as
\begin{equation}
R_{\rm astro}(z) \propto {\rm SFR}(z),
\end{equation}
potentially convolved with a delay-time distribution.  The cosmic star formation rate density rises from $z=0$ to $z\sim 2$ by roughly an order of magnitude before declining at higher redshift.

These distinct scalings lead to qualitatively different predictions for the redshift distribution of detected events.  In a mixture model,
\begin{equation}
R_{\rm tot}(z) =
f_{\rm WH}\, R_{\rm WH}(z)
+
(1-f_{\rm WH})\, R_{\rm astro}(z),
\label{eq:mixture_model}
\end{equation}
future statistically complete redshift samples can directly constrain $f_{\rm WH}$ by comparing the observed redshift distribution to dark-matter-like versus SFR-like evolution.

\subsubsection*{Spatial offsets within hosts}

A further diagnostic is the projected offset of FRBs from host-galaxy light.  White-hole events tracing the dark matter halo should follow a broader profile than the stellar light distribution, whereas magnetar progenitors are expected to correlate with star-forming regions and hence with ultraviolet or H$\alpha$ emission.  Current samples show several FRBs located within or near regions of active star formation, although the statistics remain limited \cite{Heintz_2020_hosts}.  A systematic offset analysis with larger samples could provide a decisive test of a dark-matter-tracing subpopulation.

\subsubsection*{Summary}

Host-galaxy demographics and redshift evolution currently favor progenitor channels linked to stellar processes.  A subdominant white-hole contribution is not excluded, but a dominant dark-matter-tracing population would require a redshift distribution and spatial-offset pattern inconsistent with existing data.  As localized samples expand and selection biases are better controlled, the mixture model of Eq.~\eqref{eq:mixture_model} will provide one of the most robust observational constraints on $f_{\rm WH}$.

\subsection{Synthesis: observational viability of a white-hole subpopulation}
\label{sec:obs_summary}

We now summarize the observational implications of the preceding subsections for a black-to-white-hole tunneling interpretation of fast radio bursts.

\paragraph{Rate normalization.}
The observed FRB volumetric rate, Eq.~\eqref{eq:frb_rate_obs}, sets the overall normalization.  Repetition statistics and burst morphology restrict the white-hole channel to a subset of apparently non-repeating, morphologically simple events, motivating a conservative working range of Eq.~\eqref{eq:fwh_prior}. Significantly larger fractions are disfavored unless a substantial population of intrinsically non-repeating FRBs remains hidden within present samples.

\paragraph{Spectral constraints.}
Under the assumption that the characteristic emission scale is set by the Schwarzschild radius, the observed FRB frequency range implies PBH masses $M\sim 10^{28}$--$10^{30}$\,g (Eq.~\eqref{eq:mass_range_frb}), consistent with the mass--frequency relation of Eq.~\eqref{eq:mass_freq_numeric} for $\kappa\sim 1$.  A monochromatic PBH mass function would predict clustering of spectral peak frequencies, which is not clearly observed; a broader mass function remains viable but requires detailed modeling of intrinsic spectral distributions and propagation effects.

\paragraph{Gamma-ray, gravitational-wave, and multiwavelength tests.}
For the relevant mass range, the expected prompt gamma-ray fluence of cosmological events lies several orders of magnitude below current detector sensitivities.  Prompt gravitational radiation from the transition itself falls in the GHz band, far above the reach of existing and proposed detectors, while gravitational waves from PBH binary mergers offer a complementary but currently insensitive probe of the requisite progenitor population.  Diffuse gamma-ray background constraints are likewise weak unless both $f_{\rm WH}$ and the gamma-ray efficiency approach unity.

\paragraph{Host demographics and redshift evolution.}
Observed FRB hosts and redshift distributions are broadly consistent with progenitor channels linked to star formation.  A white-hole population tracing the dark matter density would exhibit weaker intrinsic redshift evolution and a different spatial-offset distribution within hosts; current data favor SFR-like scaling, though statistical uncertainties remain substantial.

\medskip

Taken together, existing observations do not rule out a subdominant white-hole contribution to the FRB population, but they strongly disfavor a dominant origin.  A consistent interpretation is one in which white-hole events contribute at the level described by the mixture model of Eq.~\eqref{eq:mixture_model}, with $f_{\rm WH}$ at most at the few-percent to $\sim 10\%$ level.

Future progress will be driven by statistically complete redshift samples constraining $R(z)$, deep monitoring of apparently non-repeating bursts, systematic spatial-offset studies within host galaxies, and improved constraints on high-energy and gravitational-wave counterparts.  These measurements will either tighten the allowed parameter space of white-hole models or reveal distinctive signatures of a dark-matter-tracing transient population.


\section{Discussion and Conclusions}
\label{sec:conclusions}

Before summarizing our results, we collect in Table~\ref{tab:robustness} the key assumptions entering the analysis, distinguishing structural ingredients that are derived or follow from well-established physics from model-dependent ingredients whose relaxation could qualitatively change specific conclusions.  The central rate calculation (Sec.~\ref{sec:rate}) depends only on the structural ingredients; the observational discussion of Sec.~\ref{sec:obs_gamma} additionally invokes the model-dependent ones.

\begin{table}[t]
\centering
\small
\caption{Assumptions entering the analysis, where each is used, the effect of relaxing it, and whether the qualitative conclusion changes.}
\label{tab:robustness}
\begin{tabular}{p{3.2cm} p{2.0cm} p{4.5cm} p{2.8cm}}
\hline\hline
\textbf{Assumption} & \textbf{Where used} & \textbf{Effect if relaxed} & \textbf{Qualitative change?} \\
\hline
\multicolumn{4}{l}{\emph{Structural (derived / well-established):}} \\[2pt]
$\tau_{\rm WH}=\alpha(M/M_{\rm Pl})^2 t_{\rm Pl}$ & Secs.~2, 5 & Different $M$-scaling shifts ridge location & Ridge structure persists \\
Hawking evaporation $\propto M^3$ & Sec.~5 & Standard result; not relaxed & --- \\
$f_{\rm PBH,max}(M)$ bounds & Sec.~5 & Weaker bounds widen viable wedge & No: still fine-tuned \\
Monochromatic $\to$ lognormal MF & Sec.~5.1 & Enhancement $\lesssim\exp(\sigma^2/2)\sim\mathcal{O}(1)$ & No \\
Cosmological depletion & Sec.~5 & Removing it overestimates rate & No \\[4pt]
\multicolumn{4}{l}{\emph{Model-dependent (assumed / uncertain):}} \\[2pt]
$\lambda_{\rm pk}\sim\kappa R_s$ & Sec.~4.3 & Decouples FRB band from $M$ & Rate calc.\ unaffected; spectral ID model-dep. \\
$\epsilon_{\rm radio}\sim 10^{-7}$ & Sec.~3 & $D_{\rm max}\propto\epsilon^{1/2}$; factor $\sim$3 per decade & Scaling only \\
$\epsilon_{\rm HE}\sim 10^{-4}$ & Secs.~3, 6 & Shifts $\gamma$-ray horizon & No: already undetectable \\
Competitive evolution & Sec.~5 & Switch to sequential (memory burden) & Yes: collapses ridge to strip \\
Sequential (memory burden) & Sec.~\ref{sec:memory_burden} & Switch to competitive & Yes: restores broad ridge \\
$\alpha\sim 0.01$--$1$ (canonical) & Secs.~5, 7 & Larger $\alpha$ shifts ridge to higher $M$ & Ridge persists; $\gamma$-ray limits tighten \\
\hline\hline
\end{tabular}
\end{table}

In this work we have performed a systematic evaluation of the present-day and cosmological volumetric rate of primordial black hole (PBH) black-to-white-hole tunneling events.  Our treatment incorporates:
(i)~the competition between Hawking evaporation and quantum tunneling,
(ii)~cosmological depletion through the survival factor $\exp(-t/\tau_{\rm eff})$,
(iii)~realistic mass-dependent upper bounds on $f_{\rm PBH}$,
(iv)~cosmological redshift evolution of the rate density,
and (v)~the alternative memory-burden scenario enforcing sequential evolution.

The central structural result is that \emph{the burst rate is sharply peaked along the locus $\tau_{\rm eff}(M)\simeq t_0$}.  This condition defines a narrow ridge in the $(M,\alpha)$ plane (Figs.~\ref{fig:RateContour} and~\ref{fig:RateContour2}), separating an exponentially depleted region ($\tau_{\rm eff}\ll t_0$) from a long-lived region ($\tau_{\rm eff}\gg t_0$) in which the instantaneous decay probability is negligible.  In the benchmark $f_{\rm PBH}=1$ scan (Fig.~\ref{fig:RateContour}), parameter space is excluded in both directions: by overproduction in the upper-left region and by underproduction elsewhere.  Once realistic abundance constraints are imposed (Fig.~\ref{fig:RateContour2}), overproduction is largely eliminated and underproduction becomes the dominant failure mode, compressing the FRB-compatible region to a narrow wedge.

The observed FRB volumetric rate, Eq.~\eqref{eq:frb_rate_obs} intersects this ridge only in restricted regions of parameter space.  The key question is whether the canonical Planck-star range $10^{-2}\lesssim\alpha\lesssim 1$ permits FRB-level burst rates.  Our results show that it can---but only in two highly constrained, qualitatively distinct, and assumption-dependent regions.

\subsection*{Region~I: Low-mass, near-evaporation boundary}

The first viable region lies near the evaporation boundary, at masses $M\sim 10^{11}$--$10^{12}$\,kg, where Hawking evaporation and tunneling become comparable.  For $\alpha\sim 0.01$--$1$, the tunneling lifetime (Eq.~\eqref{eq:intro_tau}) becomes comparable to $t_0$ only for sufficiently small masses that lie close to the evaporation threshold; consequently only a narrow interval survives depletion.

In this region the rate can reach FRB levels if $f_{\rm PBH}$ saturates its observational upper bound, but the allowed window in $M$ is extremely narrow: slightly smaller masses are exponentially depleted and slightly larger masses have $\tau_{\rm WH}\gg t_0$.  Diffuse $\gamma$-ray limits are marginally satisfied because the mass scale is low and the total energy injection $R\,\epsilon_\gamma\, Mc^2$ remains small.  This region corresponds to the ``resonant'' condition $\tau_{\rm eff}\simeq t_0$ occurring at the evaporation boundary.  It is fine-tuned but not excluded.

\subsection*{Region~II: Sequential (memory-burden) window}

In the memory-burden scenario (Sec.~\ref{sec:memory_burden}), evaporation proceeds first down to $M_* = M/2$, after which tunneling is allowed.  This introduces a sharp threshold,
\begin{equation}
\tau_{\rm half}(M) \lesssim t_0
\quad\Rightarrow\quad
M \lesssim M_c \sim 10^{11\text{--}12}\,\mathrm{kg},
\end{equation}
and for $\alpha$ between $0.01$ and $1$ there exists a second narrow window in which $\tau_{\rm WH}(M_*)\sim t_0 - \tau_{\rm half}(M)$.  This produces FRB-level rates in a small wedge immediately below $M_c$.  In contrast to the continuous-competition case, the diagonal ridge collapses to a sharply bounded strip; the allowed region is smaller but nonzero.

An important qualitative difference is that, because early-time tunneling is forbidden, diffuse $\gamma$-ray constraints are significantly weakened relative to the no-memory case.  The viable parameter space is therefore controlled primarily by late-time burst statistics rather than cumulative energy injection.

Outside these two regions, $\alpha\sim 0.01$--$1$ fails to produce observable rates.  For $M\gg 10^{12}$\,kg the tunneling lifetime exceeds $t_0$ and the rate is suppressed as $1/\tau$; for $M\ll 10^{11}$\,kg the population is exponentially depleted.  Broad lognormal mass functions do not rescue the rate, since the integrand $R(M)$ is intrinsically narrow ($\Delta\ln M\sim 1$), and including realistic $f_{\rm PBH,max}(M)$ bounds further compresses the viable region.  Thus, the canonical Planck-star expectation $\alpha\sim\mathcal{O}(1)$ does not generically predict FRB-level event rates; it does so only under highly specific mass configurations.

\subsection*{Observational constraints and multimessenger outlook}

The observational analysis of Sec.~\ref{sec:obs_gamma} reinforces this picture from several independent directions.

\paragraph{Spectral consistency.}
If the benchmark mapping $\lambda_{\rm pk}\sim R_s$ holds (Sec.~\ref{sec:obs_spectrum}), the observed FRB frequency range $\nu\sim 0.1$--$8$\,GHz implies PBH masses $M\sim 10^{28}$--$10^{30}$\,g (Eq.~\eqref{eq:mass_range_frb}), i.e.\ Earth-to-Jupiter mass scales for $\kappa\sim 1$.  This mass window lies far above the evaporation boundary; once realistic abundance constraints are applied (Fig.~\ref{fig:RateContour2}), the rate at these masses drops well below FRB levels.  However, as discussed in Sec.~\ref{sec:obs_spectrum}, if secondary emission mechanisms shift the characteristic frequency downward, the spectral constraint is relaxed and the mass scale is no longer directly tied to the FRB band.  The rate calculation itself is unaffected by this model-dependent uncertainty.

\paragraph{Gamma-ray and diffuse background.}
For cosmological events at the mass scales implied by GHz emission, the expected prompt gamma-ray fluence lies approximately five orders of magnitude below current \textit{Fermi}/GBM and \textit{Swift}/BAT sensitivities (Sec.~\ref{sec:obs_gamma}).  Non-detections of high-energy counterparts are therefore consistent with the white-hole scenario and do not constrain it at present.  Diffuse $\gamma$-ray background limits are similarly weak unless both the white-hole fraction $f_{\rm WH}$ and the gamma-ray efficiency $\epsilon_\gamma$ approach unity, a combination already disfavored by the repetition-statistics bound of Eq.~\eqref{eq:fwh_prior}.

\paragraph{Gravitational-wave probes.}
Prompt GW emission from the transition itself falls at $\sim$GHz frequencies (Sec.~\ref{sec:obs_gw}), far above the reach of all current and proposed detectors.  Gravitational waves from PBH binary mergers at ultra-high frequencies ($10^4$--$10^6$\,Hz) offer a complementary but currently insensitive probe of the requisite progenitor population~\cite{Bird_2016_PBHmerger,Sasaki_2016_PBHmerger,MAGE_2025}.

\paragraph{Host-galaxy demographics and redshift evolution.}
Observed FRB hosts and redshift distributions are broadly consistent with progenitor channels tracing star formation rather than the dark matter density.  A white-hole population would exhibit weaker intrinsic redshift evolution and a broader spatial-offset distribution within hosts.  Current data favor SFR-like scaling (Sec.~\ref{sec:obs_hosts}), though statistical uncertainties remain substantial.  As localized samples expand and selection effects are better controlled, the mixture model of Eq.~\eqref{eq:mixture_model} will provide one of the most direct constraints on $f_{\rm WH}$.

\subsection*{Implications for quantum gravity}

The requirement $\tau_{\rm WH}(M)\sim t_0$ at $M\sim 10^{11\text{--}12}$\,kg provides a concrete phenomenological target for quantum gravity calculations of the tunneling amplitude.  If future non-perturbative computations---whether from spinfoam amplitudes, effective path integrals, or microstate overlap calculations---yield $\alpha\sim 0.1$--$1$, then PBH tunneling remains marginally viable as a subpopulation contributor to FRBs, provided PBHs saturate their abundance bounds in the critical mass window.

Conversely, if $\alpha$ is shown to be significantly smaller, the mechanism is ruled out as an FRB progenitor.  If $\alpha$ is significantly larger, the FRB-compatible ridge shifts to higher masses but rapidly encounters both diffuse $\gamma$-ray constraints and the requirement to reconcile the emission frequency with the Schwarzschild scale.

\subsection*{Summary of conclusions}

Our main conclusions are:
\begin{enumerate}
\item The PBH tunneling burst rate is maximized along the condition $\tau_{\rm eff}\simeq t_0$, defining a narrow ridge in the $(M,\alpha)$ plane.

\item For $0.01\le\alpha\le 1$, FRB-level rates arise only in two narrow regions:
    (i)~a low-mass window near the evaporation boundary, and
    (ii)~a sequential (memory-burden) wedge below the critical mass $M_c\sim 10^{11\text{--}12}$\,kg.
    Outside these regions the predicted rate falls many orders of magnitude below observed FRB densities.

\item Multiwavelength and multimessenger observations---prompt gamma-ray non-detections, diffuse background limits, host-galaxy demographics, and redshift evolution---are all consistent with a subdominant white-hole contribution but strongly disfavor a dominant origin.

\item The viable parameter space is not broad but highly fine-tuned, requiring PBHs to saturate their abundance upper bounds in a narrow mass window where the tunneling lifetime is comparable to the age of the Universe.
\end{enumerate}

\noindent The overarching conclusion of this work is that the PBH tunneling rate calculation does not generically support FRB-level event densities: any viable interpretation requires narrow, fine-tuned, and strongly assumption-dependent corners of parameter space.  Planck-star phenomenology remains a logically possible but tightly constrained explanation for a small fraction of FRBs.  Future advances in quantum gravity calculations of the tunneling amplitude, improved PBH abundance constraints across the asteroid-to-planetary mass range, and statistically complete high-redshift FRB samples will collectively determine whether this scenario can be sharpened into a testable prediction or is definitively excluded.

\section*{Acknowledgments}

This work was supported in part by the U.S. Department of Energy, Office of Science, Office of High Energy Physics under grant Contract Number DE-SC010107.

\bibliographystyle{JHEP}

\bibliography{biblio.bib}

\end{document}